\documentclass[%
aip,
 amsmath,amssymb,
 reprint,%
]{revtex4-1}

\usepackage{graphicx}
\usepackage{dcolumn}
\usepackage{bm}

\usepackage[T1]{fontenc}
\usepackage{mathptmx}
\usepackage[colorlinks,linkcolor=red, anchorcolor=blue, citecolor=blue]{hyperref}

\begin{document}

\preprint{AIP/123-QED}

\title{Designing Heat Transfer Pathways for Advanced Thermoregulatory Textiles}

\author{Xiaohua Lan}
\thanks{These two authors contributed equally to this work.}
\affiliation{Center for Phononics and Thermal Energy Science, China-EU Joint Lab on Nanophononics, Shanghai Key Laboratory of
Special Artificial Microstructure Materials and Technology, School of Physics Science and Engineering, Tongji University, Shanghai
200092, China}

\author{Yi Wang}
\thanks{These two authors contributed equally to this work.}
\affiliation{Center for Phononics and Thermal Energy Science, China-EU Joint Lab on Nanophononics, Shanghai Key Laboratory of
Special Artificial Microstructure Materials and Technology, School of Physics Science and Engineering, Tongji University, Shanghai
200092, China}

\author{Jiebin Peng}
\affiliation{Center for Phononics and Thermal Energy Science, China-EU Joint Lab on Nanophononics, Shanghai Key Laboratory of
Special Artificial Microstructure Materials and Technology, School of Physics Science and Engineering, Tongji University, Shanghai
200092, China}
\author{Yang Si}
\affiliation{Innovation Center for Textile Science and Technology, Donghua University, Shanghai 200051, China}
\author{Jie Ren}
\email{xonics@tongji.edu.cn}
\affiliation{Center for Phononics and Thermal Energy Science, China-EU Joint Lab on Nanophononics, Shanghai Key Laboratory of
Special Artificial Microstructure Materials and Technology, School of Physics Science and Engineering, Tongji University, Shanghai
200092, China}
\author{Bin Ding}
\email{binding@dhu.edu.cn}
\affiliation{Innovation Center for Textile Science and Technology, Donghua University, Shanghai 200051, China}
\author{Baowen Li}
\email{baowen.li@colorado.edu}
\affiliation{Paul M Rady Department of Mechanical Engineering and Department of Physics, University of Colorado Boulder, CO80309, USA.}

\date{\today}

\begin{abstract}
Thermal comfort of textiles plays an indispensable role in the process of human civilization.  Advanced textile for personal thermal management shapes body microclimates by merely regulating heat transfer between the skin and local ambient without wasting excess energy.
Therefore, numerous efforts have recently been devoted to the development of advanced thermoregulatory textiles. In this review, we provide a unified perspective on those state-of-the-art efforts by emphasizing the design of diverse heat transfer pathways. We focus on engineering certain physical quantities to tailor the heat transfer pathways, such as thermal emittance/absorptance, reflectance and transmittance in near-infrared and mid-infrared radiation, as well as thermal conductance in conduction. Tuning those heat transfer pathways can achieve different functionalities for personal thermal management, such as passive cooling, warming, or even dual-mode (cooling-warming), either statically switching or dynamically adapting. Finally, we point out the challenges and opportunities in this emerging field, including but not limited to the impact of evaporation and convection with missing blocks of  heat pathways, the bio-inspired and artificial-intelligence-guided design of advanced functional textiles.
\end{abstract}

\maketitle

\tableofcontents

\section{\label{sec1}Introduction}Textiles, an essential part of our life, provide us not only with a thermal conform, but also personal identity and aesthetics \cite{ref248}. With the development of science and technology, modern textiles are designed to explore more diverse properties for human in medical, fire-fighting, aerospace and military fields, such as bacterial resistance, flame retardant, light weight, breathability, waterproofness, durability, etc \cite{ref249}. Although these functional textiles have focused on different textile properties, adapting thermal comfort to varied ambient has always been the eternal primary function of textiles, no matter how human society evolves. As we all know, human are endotherms that maintain the constant
temperature of 37 $^\circ$C around, which is also the range of human thermal comfort.

Human thermal comfort can be perceived simply as the state that the surface temperature of the skin is converging on the normal body
temperature as much as possible.  Thermal comfort is closely associated with multiple heat transfer pathways among the skin, textile,
sun and the external environment \cite{ref11, ref12,ref73,ref74}. Importantly, the heat transfer is strongly dependent on the
external environmental temperature. In the hot summer, prodigious amounts of heat transfer from the external environment to the skin
increase the skin temperature. On the other hand, heat mainly transfers from the skin to the environment, decreasing the
temperature of skin in the cold winter.
In addition, the heat transfer rates between the environment and skin covered with different textiles obtain certain differences,
which depend on the material, microstructure and design of textiles \cite{ref14, ref13, ref15}. However, traditional textiles cannot effectively
suppress the human heat dissipation in cold conditions, and also play limited roles in improving the heat dissipation in hot conditions causing a rising skin temperature.
Consequently, numerous efforts are explosively emerging to explore the advanced thermoregulatory textiles to extend the range of comfortable ambient temperature of the human body.

Here, we divided advanced thermoregulatory textiles into three categories: passive cooling,  warming  and dual-mode (cooling-warming switch) textiles. Cooling textiles can enhance the heat transfer rate to relieve the human heat stress in hot conditions, and  warming  textiles  can  reduce  the  heat  loss  of  body to achieve the thermal insulation effect in the cold environment. The dual-mode textiles have an attractive mechanism that the heat transfer rate can adaptively change according to the environmental variation, which is beneficial to the human body's ability of acclimatizing to the changeable environment.
Different from existing few reviews summarized emerging thermoregulatory textiles from the perspectives of materials engineering \cite{ref14,ref74,ref73} and functionalities \cite{ref234}, we provide a unified perspective on emphasizing the design of diverse heat transfer pathways in advanced thermoregulatory textiles.
We focus on engineering certian thermophysical quantities to regulate the heat transfer pathways, including thermal emittance, reflectance and transmittance in thermal radiation, as well as non-radiative thermal conductance.

We do not concern about using latent heat in phase-change materials \cite{ref14,ref74,ref73} or active cooling by external powers \cite{ref73,ref234}. We mainly focus on thermal radiation and non-radiation to tailor the heat transfer pathways in textiles to achieve different functionalities for personal thermal management. This review is organized as follows: Firstly, fundamental theories are
introduced in Section \ref{sec2}, such as the concept of thermal comfort and balance, the principle of thermal radiation, and non-radiative conduction,
and heat transfer pathways in body microclimate. We then discuss the advanced thermoregulatory textiles according to different
mechanisms of heat transfer pathways for passive cooling (Section \ref{sec3}), warming (Section \ref{sec4}) and dual-mode (Section
\ref{sec5}), respectively. Finally, we summarize with illustrating the challenges and providing perspectives on the rapid development of this field (Section \ref{sec6}), including the impact of evaporation and convection with missing blocks of heat pathways, and so on.

\section{\label{sec2}Fundamentals of human body thermal regulation}
In daily life, we can adjust the ambient temperature to achieve
human thermal comfort through external equipment, such as fans, air conditioners and heaters, which will consume amounts of energy. On the other hand, studies have shown that enhancing the individual microclimate is more comfortable than improving the climate throughout the whole space due to the differences of human thermal sensation \cite{ref17}. Meanwhile, improving personal microclimate can also reduce unnecessary energy consumption. To better regulate personal microclimate,
we need to consider the concept of human thermal comfort, the mechanism of heat transfer between the clothed skin and environment, and the multiple heat transfer pathways. We will discuss these specific details in the following.
\begin{figure*}
\centering
    \includegraphics[scale=0.7]{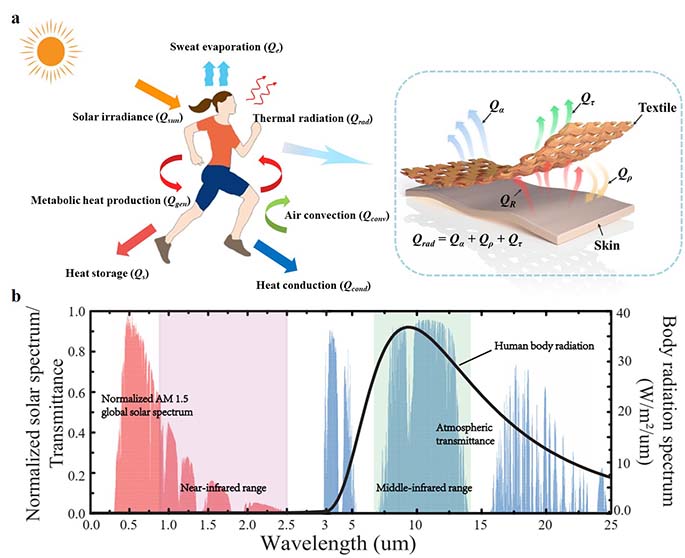}
		\caption{(a) Schematic of the heat transfer pathways between the human body and the outdoor environment. In the outdoor circumstance, the solar irradiance ($Q_{sun}$) and metabolic heat generation ($Q_{gen}$) are the precious sources of heat. In general, thermal radiation ($Q_{rad}$), heat conduction ($Q_{cond}$), air convection($Q_{conv}$), and sweat evaporation ($Q_{e}$) are regarded as primary heat dissipation pathways of human body. $Q_{s}$ is the heat storage of human body. The illustration depicts that the radiative energy of the human body ($Q_{rad}$) is reflected ($Q_{\rho}$), absorbed ($Q_{\alpha}$) and transmitted ($Q_{\tau}$) by the textile, which further indicates the significant influence of textile on the heat transfer between the skin and environment. (b) The radiation spectrum of the human body was plotted according to the black body radiation law at a temperature of 310 K (corresponding to the normal skin temperature of 37 $^\circ$C\ ). The normalized AM 1.5 global solar spectrum and the atmospheric transmittance spectrum are shown as orange and blue background map. The ranges of purple and light green regions illustrate the near-infrared (NIR) and middle-infrared (MIR) band, respectively \cite{ref233}.}
\label{Fig.1*}
\end{figure*}

\subsection{\label{sec2.2}Concept of human thermal comfort and balance}Comfort is a term created by psychologists, but its connection to the physiological basis is still vague \cite{ref20}. Comfort is the same as ``optimal temperature'', but even
the term ``optimal temperature'' is defined differently by physiologists and behavioral scientists \cite{ref20}. According to the
American Society of Heating, Refrigerating and Air Conditioning Engineers (ASHRAE) Standard, human thermal comfort is defined as a state of mind that
is satisfactory to the thermal environment \cite{ref18}. However, each of us has obvious differences on the satisfaction with the same thermal environment, which depends on the physical activity, metabolism level and temperature sensitivity of the body.
In any case, thermal comfort is close to the thermal balance of the human body.

The body achieves a balance between its heat gain and loss, maintaining a constant body temperature of 37 $^\circ$C. According to Fig. \ref{Fig.1*}a, this
balance can be expressed by the body heat balance equation, which is shown in the following:

\begin{equation}
\label{E3}
Q_{sun}+Q_{gen}=Q_{rad}+Q_{cond}+Q_{conv}+Q_{e}+Q_{s}
\end{equation}
where $Q_{sun}$ is heat absorption from the sun, $Q_{gen}$ is the metabolic heat generation, $Q_{rad}$, $Q_{cond}$, $Q_{conv}$, and $Q_{e}$ are the radiative heat transferred to the environment by radiation, conduction, convection, and sweat evaporation. $Q_{s}$ is the body heat storage.

Although Eq. (\ref{E3}) can be varied, they involve similar
concepts, such as human body heat generation, transfer and storage. In outdoor situation, human body will
absorb heat from the solar due to the almost perfect absorption of skin with an excellent emittance of 0.98 while the term
can be ignored under the indoor condition \cite{ref38}. Part of the metabolic heat generation is used to maintain the body for daily activities,
and the rest is released as heat. The body can release heat to the environment by means of thermal radiation, heat conduction, air convection and sweat evaporation (Fig. \ref{Fig.1*}a). In conclusion, heat generation and dissipation of the body will eventually produce a body heat storage.

The $Q_{sun}$ is a extra heat gain of human body under the outdoor condition, which can be expressed as
\begin{equation}
\label{E10}
Q_{sun}=\int_{0.3}^{4}\ I_{AM1.5}(\lambda)\cdot \alpha(\lambda)d\lambda
\end{equation}
where $I_{AM1.5}(\lambda)$ represents the Air Mass (AM) 1.5 global solar irradiance spectrum, $\lambda$ is the radiative wavelength.
$\alpha(\lambda)$ is the absorptance of the textile in the range of 0.3 $\mu m$ - 4 $\mu m$.

The value of $Q_{gen}$ is associated with the activity level of the body. Specifically, the $Q_{gen}$ is a primary heat gain of the human body. The metabolic heat generation possesses many ways, which include blood circulation,
skeletal muscle, heat produced by the liver, and chemical heat produced by oxidation of nutrients absorbed inside the body
\cite{ref7}.

In addition, more than 50\% of the heat of the human body is released to the environment in the form of thermal radiation
\cite{ref28}, and human thermal radiation is regarded as nearly perfect blackbody radiation in the middle infrared band \cite{ref26, ref27}. As
shown in Fig. \ref{Fig.1*}b, the human body emits electromagnetic waves in the range of 7 to 15 $\mu m$ at the skin temperature of 37
$^\circ$C\ , which has an emission peak at 9 $\mu m$ roughly. Obviously, the thermal radiation range of the human body overlaps with
the atmospheric window. Thus, the radiative heat can transfer directly to the cold outer space (3K) through the atmospheric
window (atmosphere has a minimum absorption for radiative wavelength of objects between 8-13 $\mu m$). The heat loss of human body through thermal radiation can be calculated as:
\begin{equation}
\label{E11}
Q_{rad}=\int_{0}^{\infty}\ I_{BB}(T, \lambda)\cdot \epsilon(\lambda)d\lambda
\end{equation}
where, $\epsilon(\lambda)$ is the emittance of the skin or textile, $I_{BB}(T, \lambda)$ is spectral emissive power of a blackbody and T is the temperature. Notably, the radiative heat transfer between body and environment can be influenced by the optical properties ($\alpha, \rho$  and
$\tau$) of textiles in the middle infrared band. However, traditional textiles enhance human thermal comfort rarely using solar irradiance and thermal radiation.

In the outdoor conditions, the solar energy exerts an essential influence on the heat transfer of the human body. Solar spectrum is mainly composed of ultraviolet light (280-400 nm), visible light (400-700 nm), and near  infrared radiation (NIR) light (700-2500 nm). Specifically, the energy distributions of ultraviolet, visible, and NIR light are approximately 2 $\%$, 47 $\%$, and 52 $\%$, respectively \cite{ref233}. However, most of researchers major discussed the heat absorption of the human body in the NIR band as it is hard to design the advanced thermoregulatory textiles by adjusting the optical properties of materials in the whole solar spectrum \cite{ref156,ref262,ref174}. Therefore, we also focus on considering the NIR absorption of the human body and reviewing in this field. In addition, human thermal radiation is concentrated on the middle infrared radiation (MIR) within the range of 7-14 $\mu m$. As a result, the method of regulating the human heat transfer and designing the structure of textiles to adjust the optical properties in the MIR and NIR ranges have already become a hot topic.

Thermal conduction ($Q_{cond}$) depends on the temperature gradient and thermal conductance of the medium. The thermal conductance of textile also exerts a substantial impact on the heat transfer between skin and textile, indicating the potential method to regulate the heat transfer of body through modulating the thermal conductance ($G$). Moreover, air convection ($Q_{conv}$) also has an influence on the heat transfer between skin, textile, and environment in different degrees that can be well illustrated by $\gamma$ depending on the personal activity, the intrinsic properties of textile, and specific environment. The heat transfer rate of clothed skin-environment system is ultimately determined by the effective heat-transfer coefficient consisting of the conduction and convection coefficient. More details about the quantitative analysis of heat loss caused by heat conduction and convection, shown is in section \ref{sec2.3}.

Notably, sweat evaporation ($Q_{e}$) is the unique cooling mechanism of human body in the case of the skin with higher humidity. The sweat will transfer from the skin to the outer surface of textile induced by the capillary action of textile. Then, the sweat on the surface of textile is evaporated due to the heat conducted through skin, which will efficiently enhance the heat transfer of human body as the water vaporization has the high latent heat \cite{ref250}. Obviously, the extra cooling power contributed by the sweat evaporation is related to the wicking performance and effective heat-transfer coefficient of textile. Therefore, researchers also have developed some advanced thermoregulatory textiles by adjusting the effective heat-transfer coefficient and moisture permeability of textiles.

The last item of the Eq. (\ref{E3}) is the heat storage of human body ($Q_{s}$) due to the difference between the heat gain and loss. If heat storage is 0, it means that the heat generation and dissipation of the body are in equilibrium, indicating that the
body is in a state of thermal comfort. If the heat storage is greater than 0, the body gets a net heat generation, which results in increase of body temperature gradually. On the contrary, the heat storage is less than 0 causing the drop of body temperature \cite{ref18}. Deviating from the normal body temperature will exert a significant influence on thermal comfort and can even induce sickness.

\subsection{\label{sec2.1}Principle of thermal radiation and conduction}
{\bf 1. Thermal radiation}

In human body thermal regulation, when the radiative energy ($Q_{rad}$) propagates to the real surfaces with roughness or texture, it can be absorbed ($Q_{\alpha}$), reflected ($Q_{\rho}$) and transmitted ($Q_{\tau}$) by the object. Various parts of radiative energy satify the law of energy conservation (Fig. \ref{Fig.1*}a).
\begin{align} \label{Q_fun}
  Q_{rad}=Q_{\alpha}+Q_{\rho}+Q_{\tau}
\end{align}
The proportion of partial energy to total energy can be described respectively as the absorptance $\alpha$, reflectance $\rho$ and transmittance $\tau$. Eq.~\eqref{Q_fun} with temperature T can be normalized as below:
\begin{align} \label{Q_norm}
  {\alpha}(T)+{\rho}(T)+{\tau}(T)=1
\end{align}

{\it Emittance, absorptance and Kirchhoff's Law--}For real object, the ratio between the emissive energy flux of the object and that of a blackbody with temperature T can define the emittance. The total emittance can be calculated by integration of the
spectral emittance in the whole wavelength region weighted by the blackbody distribution:
\begin{align} \label{total_em}
    &\epsilon_{tot}(T) = \frac{\int_{0}^{\infty} I_{BB}(\lambda,T) \epsilon^h(\lambda) d\lambda }{\sigma_{SB} T^4}
\end{align}
where $\epsilon_{tot}(T)$ and $\epsilon^h(\lambda)$ are the total emittance and hemispherical spectral
emittance. $\sigma_{SB}$ is
the Stefan-Boltzmann constant. $I_{BB}(\lambda,T)$ is the spectral emissive power of a blackbody with temperature T (Planck's Law).
\begin{equation}
\label{E1}
I_{BB}(T, \lambda)=\frac{2hc^{2}}{\lambda^{5}}\frac{1}{e^{\frac{hc}{\lambda k_{B}T}}-1}
\end{equation}
where $h$ is Planks's constant, $c$ is the light velocity, $k_B$ is Boltzmann constant.

For thermal radiation, Kirchhoff's Law can be derived from the principle of detailed balance, which illustrates the equality
between absorptance and emittance:
\begin{align}
\alpha(T)=\alpha_{tot} (T) = \epsilon_{tot} (T)
\end{align}
where $\alpha_{tot}$ is the total absorptance. Under Lorentz
reciprocity conditions, the Kirchhoff's Law can be proven based on the fluctuation-dissipation theorem. Past investigations have
indicated that the human skin can perform like a blackbody with a near-unity emittance in the IR wavelength range \cite{ref38}. For
convenience, we assume that the all surfaces in human body thermal regulation are diffuse-gray so that Kirchhoff's Law is valid
for emittance and absorptance.

{\it Reflectance and transmittance--}Under the considerations of local thermal equilibrium and reciprocal materials, the total reflectance can be defined by the fraction between the reflected and incident energy. Assuming that the spectral intensity of radiation with temperature T is $I(\lambda,T)$, the general form of total reflectance can be described as below:
\begin{align} \label{total_ref}
      \rho(T)=\rho_{tot}(T)=\frac{\int_{0}^{\infty} I(\lambda,T) \rho(\lambda) d\lambda }{\int_{0}^{\infty} I(\lambda,T) d\lambda }
\end{align}
where $\rho(\lambda)$ is hemispherical reflectance. However, the reflectance of radiation by real surfaces is complicated. For surfaces with random
roughness or texture, the specular, off-specular, and diffuse reflection play different roles in the total reflectance. However, when
the periodic structures exist in the surface, diffraction and coherent effects can become crucial, and several peaks can appear in the
spectral function of reflectance. Usually, people use bidirectional reflectance distribution function to describe the
reflectance of a rough surface at a given wavelength or frequency. And it can be defined as the reflected light intensity divided
by the incident light flux at the surface \cite{ref245}. We only give the general relation between hemispherical reflectance and bidirectional reflectance distribution function below:
\begin{align}\label{BRDF}
 \rho(\lambda) = \int_{0}^{2\pi} \int_{0}^{2\pi} f_r(\lambda,\theta_i,\phi_i,\theta_r,\phi_r) \cos \theta_i \cos \theta_r d\Omega_i d\Omega_r
\end{align}
where $f_r(\lambda,\theta_i,\phi_i,\theta_r,\phi_r) $ and $[\theta_{i(r)},\phi_{i(r)}]$ are bidirectional reflectance distribution function and directions of incident(reflected) light, respectively. $d\Omega_{i(r)}$ is the solid angle for incident(reflected) light.

In general, the bidirectional reflectance distribution function of a real surface can be calculated by solving the Maxwell equations if the surface roughness is well measured.
There are several numerical and approximate methods to calculate the actual bidirectional reflectance distribution function. For example, the boundary integral
method, the Rayleigh-Rice perturbation theory, the Kirchhoff approximation and the geometric optics approximation. These
approximations are valid only for certain conditions of roughness and wavelength. More information for the discussion on the
roughness parameters as well as the instruments used for surface characterization can be found in reference \cite{ref246}.

Similarly, the total transmittance can be defined by the fraction between the transmitted and incident energy for broadband thermal radiation and the general form can be showed as below:
\begin{align}\label{EQ_tran}
  \tau(T)=\tau_{tot}(T) = \frac{\int_{0}^{\infty} I(\lambda,T) \tau(\lambda) d\lambda }{\int_{0}^{\infty} I(\lambda,T) d\lambda }
\end{align}
where $\tau(\lambda)$ is the hemispherical or spectral transmittance. According to the fluctuation-dissipation theorem, the thermal radiation is originated from the random charge fluctuations. If we evaluate the incident thermal energy using Planck's distribution in Eq.~\eqref{total_ref} and Eq.~\eqref{EQ_tran}, the Eq.~\eqref{total_em}-like formula can be given for total refectance and transmittance.

In most studies for human body thermal regulation, the IR reflectance $\rho$ and transmittance $\tau$ can be measured using a FTIR
spectrometer. The direct calculations for emittance $\epsilon$ can be accomplished by considering Kirchhoff's Law, along with
absorptance $\alpha$ inside the materials.

{\bf 2. Thermal conduction and convection}

Thermal conduction is a common heat transfer pathway between two contact objects with the temperature gradient through the collision and motion of microscopic particles. Fourier's Law indicates the specific relationship between the temperature gradient and the heat flux per unit area in the process of heat transfer:
\begin{align}
\label{E2}
\dot{Q}_{cond}=-\kappa \nabla T=G \Delta T,\;\;\;\;\;\; G=\kappa/t,
\end{align}
where $G$ is defined as thermal conductance that is related to the thermal conductivity ($\kappa$) and thickness ($t$) of textile. $\Delta T$ is the temperature difference, related to the temperature gradient $\nabla T=-\Delta T/t$. The thermal conductance performs an essential role in modulating the heat transfer rate between the human body and textiles with a temperature difference, considering both the effects of thermal conductivity and thickness of the textiles.

Thermal convection is the process of heat transfer for the fluid system with the temperature difference, which depends on the motion of the fluid. Thermal convection is divided into the natural and forced convection based on the external sources. In general, two types of convection exist in the heat transfer of clothed skin-environment. Thermal convection is governed by Newton's Law of cooling, which can be expressed as:
\begin{equation}
\label{E12}
\dot{Q}_{conv}=\gamma (T_{h}-T_{l}),
\end{equation}
where $\gamma$ is the convective heat-transfer coefficient, $T_{h}$ is the temperature of object and $T_{l}$ is the temperature of fluid, respectively. $\gamma$ is related to the viscosity, density, specific heat capacity, and thermal conductivity of fluid. Specifically, the activity level of human body and air velocity exert a great influence on the heat transfer between the human body and external environment.

\subsection{\label{sec2.3}Heat transfer pathways in body microclimate}
\begin{figure}
\centering
    \includegraphics[scale=0.4]{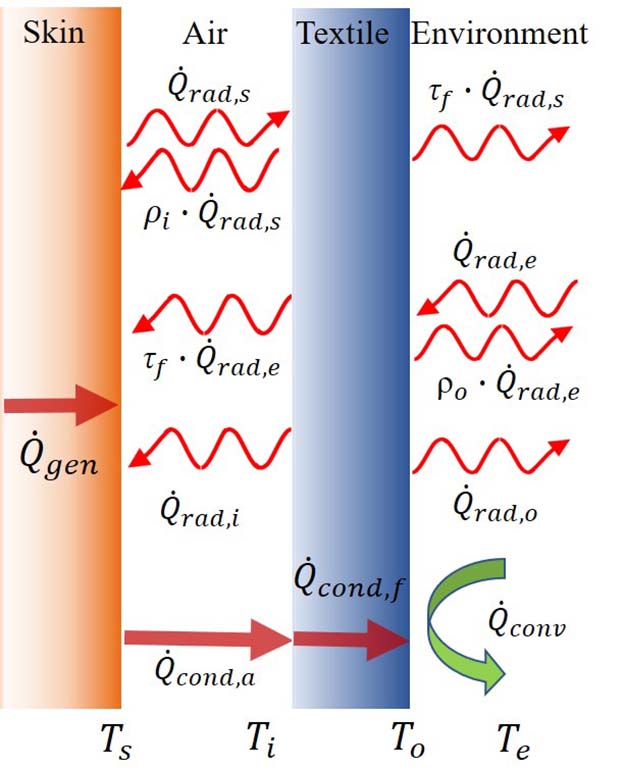}
	\caption{Schematic depicts concretely the multiple heat transfer ways between clothed skin and the environment using the steady-state heat transfer model, including thermal radiation, conduction, and convection.}
\label{Fig.2*}
\end{figure}

In the previous sections, we have discussed some mechanisms of regulating heat transfer between clothed skin and environment from a
qualitative point. Here, the one-dimensional steady-state heat transfer model (Fig. \ref{Fig.2*}) is used to analyze the heat exchange mechanism of the relatively simple situation when the human body works indoors and in a sedentary state \cite{ref23}. In this steady-state model, the body's heat source is mainly
the body's metabolic heat generation. Thermal radiation, heat conduction, air convection and sweat evaporation are the vital ways that the body dissipate heat to the environment. Since the skin surface is covered with textiles, a part of the radiative heat from the human body is reflected back to the skin surface by the textiles. Part of the heat is absorbed by the textiles, and the remaining heat passes through the textiles to the external environments. Similarly, some of the radiative heat from the atmosphere is reflected by the textiles to the atmosphere, a part of which is absorbed by the textiles. And the rest of the heat passes through the
textiles to skin. In addition, heat can be exchanged from the clothed skin to the environment by means of heat conduction, convection and sweat evaporation. Convective heat transfer is negligible because the human body is assumed to be stationary between the skin, inner, and outer surface of textile. Meanwhile, the heat dissipation of sweat evaporation is also neglected as it accelerates effectively the heat dissipation of human body in the higher relative humidity of skin.
\begin{figure*}
\centering
		\includegraphics[scale=0.485]{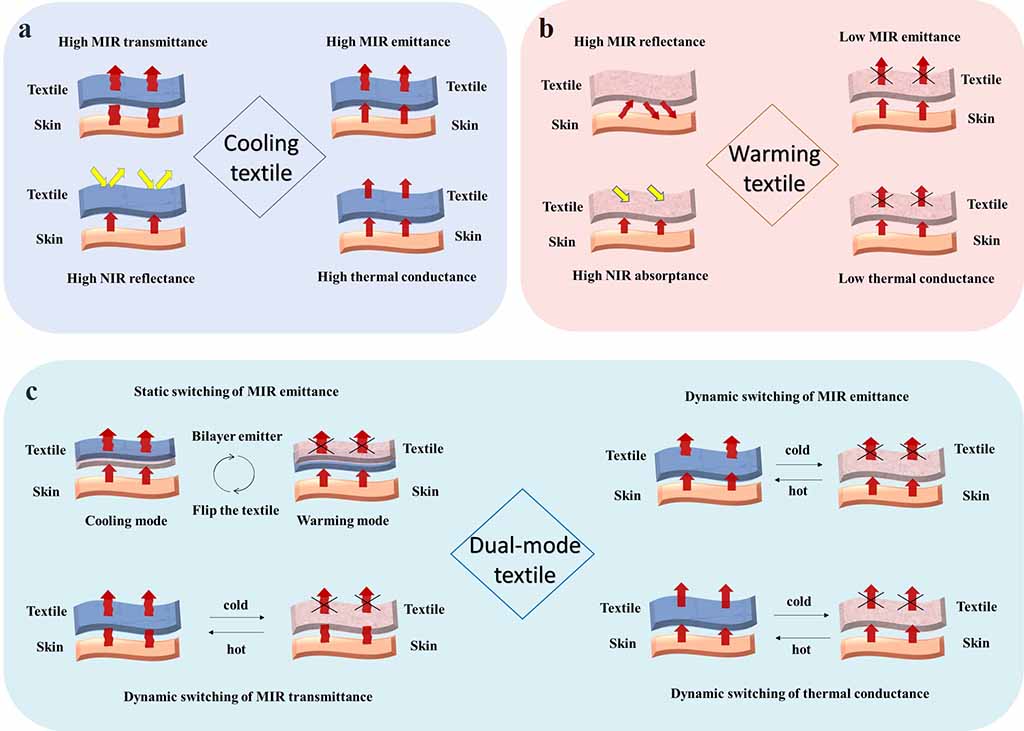}
	\caption{Different design principles of (a) cooling textiles, (b) warming textiles and (c) dual-mode textiles in the current studies based on modulating the heat transfer ways between human body and external environment. Cooling textiles (a) that accelabrate the heat dissipation can be designed by enhancing the MIR transmittance, emittance, NIR reflectance, and thermal conductance of textiles. Correspondingly, warming textiles (b) that inhibit the heat loss can be achieved through improving the MIR reflectance, NIR absorptance and reducing the MIR emittance, thermal conductance of textiles. Currently, dual-mode textiles (c) switching between cooling and warming are divided into statically dual-mode ones and dynamically dual-mode ones. Embedding the bilayer emmiter with different emittances into the textiles can switch the cooling and warming modes statically by flipping the textiles. Dynamically dual-mode textiles can switch the heat transfer of the skin based on the variable environment, which are developed via regulating the MIR emittance, transmittance, and thermal concuctance of textiles. }
\label{Fig.3*}
\end{figure*}

In the model, human thermal comfort need to meet the condition when the equality of the total heat loss rate with the total heat generation
rate. According to the principle of energy balance in the skin and textile system (Fig. \ref{Fig.2*}), we can list the following equilibrium equations
for each system of the model:

\noindent In the skin system:
\begin{equation}
\label{E4}
\dot{Q}_{gen}+\rho_{i}\cdot \dot{Q}_{rad,s}+\tau_{f}\cdot \dot{Q}_{rad,e}+\dot{Q}_{rad,i}=\dot{Q}_{rad,s}+\dot{Q}_{cond,a}
\end{equation}
In the textile system:
\begin{equation}
\label{E5}
\begin{split}
\dot{Q}_{rad,s}+\dot{Q}_{rad,e}&+\dot{Q}_{cond,a}=\rho_{i}\cdot \dot{Q}_{rad,s}+\tau_{f}\cdot \dot{Q}_{rad,e}+\dot{Q}_{rad,i}\\
&+\tau_{f}\cdot \dot{Q}_{rad,s}+\rho_{o}\cdot \dot{Q}_{rad,e}+\dot{Q}_{rad,o}+\dot{Q}_{conv}
\end{split}
\end{equation}
where $\dot{Q}_{gen}$ (the metabolic heat generation per unit area) is assumed to be 73 W/m$^{2}$ corresponding to the human body in a sedentary state \cite{ref23,ref258}, $\dot{Q}_{rad,s}$ represents the heat flux radiated from the skin, $\dot{Q}_{rad,e}$ is the heat flux radiated from the atmosphere to the human body, $\dot{Q}_{rad,i}$ is the heat flux radiated from the inner surface of the textile, $\dot{Q}_{rad,o}$ is the heat flux from the outer surface of the textile to the external environment, $\dot{Q}_{cond,a}$ represents the conduction heat flux between the skin and the clothings, $\dot{Q}_{conv}$ is the heat convection between the clothing and the external environment. $\rho_{i}$ and $\rho_{o}$ are the IR reflectance of the inner and outer textile, respectively. $\tau_{f}$ is the IR transmittance of the textile.

The heat flux of thermal radiation, conduction and convection can be solved based on Stefan-Boltzmann Law, Fourier's Law, and
Newton's Law of cooling, respectively.
\begin{align}
\label{E6}
\begin{split}
&\dot{Q}_{rad,s}=\varepsilon_{s}\sigma_{SB}{T_{s}}^{4}, \quad
\dot{Q}_{rad,e}=\varepsilon_{e}\sigma_{SB}{T_{e}}^{4} \\
&\dot{Q}_{rad,i}=\varepsilon_{i}\sigma_{SB}{T_{i}}^{4}, \quad
\dot{Q}_{rad,o}=\varepsilon_{o}\sigma_{SB}{T_{o}}^{4}
\end{split}
\\
&\dot{Q}_{cond,a}=G_{a}\cdot(T_{s}-T_{i}) \\
& \dot{Q}_{conv}=\gamma (T_{o}-T_{e})
\end{align}
where the skin temperature ($T_{s}$) is assumed to 37 $^\circ$C\ , $T_{e}$ is the ambient temperature, $T_{i}$ and $T_{o}$ represent the temperature of inner surface and outer surface textile, respectively. $\sigma_{SB}$ is the Stefan-Boltazmann constant. $\varepsilon_{s}$ and $\varepsilon_{e}$ are the MIR emittance of skin and ambient,respectively. $\varepsilon_{i}$ and $\varepsilon_{o}$ represent the MIR of emittance of the inner and outer surface of textile, respectively. $G_{a}$ is the thermal conductance of air. In this model, natural convection is regarded as the most important factor in the progress of heat convection between the clothed skin and environment. Therefore, the convective heat transfer coefficient ($\gamma$) can be estimated according to the formula of natural convective heat transfer coefficient ($ \gamma=2.38(T_{o}-T_{e})^{0.25} $) \cite{ref257}.

In addition, thermal conduction ($\dot{Q}_{cond,f}$) is the primary heat transfer mechanism of textile. Taking into account the temperature profile of textile, the inner and outer surface temperatures of the textile can be analyzed as follows \cite{ref24,ref23,ref258}:
\begin{equation}
\label{E9}
\begin{aligned}
\begin{split}
&T_{o}=\frac{1}{2G_{f}}[\dot{Q}_{rad,i}+\dot{Q}_{rad,o}-\alpha_{i}(\dot{Q}_{rad,s}+ (1-\varepsilon_{s}) \\
&(\rho_{i}\cdot \dot{Q}_{rad,s}+\tau_{f}\cdot \dot{Q}_{rad,e}+\dot{Q}_{rad,i}))- \alpha_{o}\dot{Q}_{rad,e}] \\
&-\frac{G_{a}}{G_{f}}\cdot(T_{s}-T_{i})+T_{i}
\end{split}
\end{aligned}
\end{equation}
In the above equations, we need input some fixed parameters that can be measured experimentally, such as, the optical parameters of textile ($\tau_{f}$, $\rho_{i}$, $\rho_{o}$, $\varepsilon_{i}$, $\varepsilon_{o}$, $\varepsilon_{e}$, and $\varepsilon_{s}$), and the thermal conductance of textile ($G_{f}$). Then, we can obtain the comfortable environmental temperature ($T_{e}$) of textiles with different optical and conductive properties from the simultaneous Eq. (\ref{E4})-Eq. (\ref{E9}). The heat transfer model provides the inspiration for the design of advanced thermoregulatory textiles.

Importantly, different design principles of cooling, warming and dual-mode textiles are basis of modulating the heat transfer ways between human body and external environment, as shown in Fig. \ref{Fig.3*}. Meanwhile, we focus on elaborating physical quantities to tailor the heat transfer pathways with considering thermal emittance/absorptance, reflectance and transmittance in NIR and MIR, and thermal conductance in conduction and convection.

Cooling textiles in Fig. \ref{Fig.3*}a that accelerate the heat dissipation can be designed by enhancing
the MIR transmittance, emittance, NIR reflectance, and thermal conductance of textiles, corresponding to Section \ref{sec3.1}, \ref{sec3.2}, \ref{sec3.3} and \ref{sec3.4}. Warming textiles in Fig. \ref{Fig.3*}b that inhibit the
heat loss can be achieved through improving the MIR reflectance, NIR absorptance and reducing the MIR emittance, thermal conductance
of textiles, corresponding to Section \ref{sec4.1}, \ref{sec4.2}, \ref{sec4.3} and \ref{sec4.4}. Meanwhile, dual-mode textiles in Fig. \ref{Fig.3*}c switching between cooling and warming are divided into statically dual-mode one and dynamically
dual-mode one. Embedding the bilayer emitter with different emittance into the textile can switch the cooling and warming mode statically
by flipping the textile in Section \ref{sec5.1}. Dynamically dual-mode textile can switch the heat transfer of the skin based on the changeful environment, which is
developed via regulating the MIR emittance, transmittance, and thermal conductance of textile corresponding to Section \ref{sec5.2}, \ref{sec5.3} and \ref{sec5.4}.

\begin{figure}
\centering
		\includegraphics[scale=0.423]{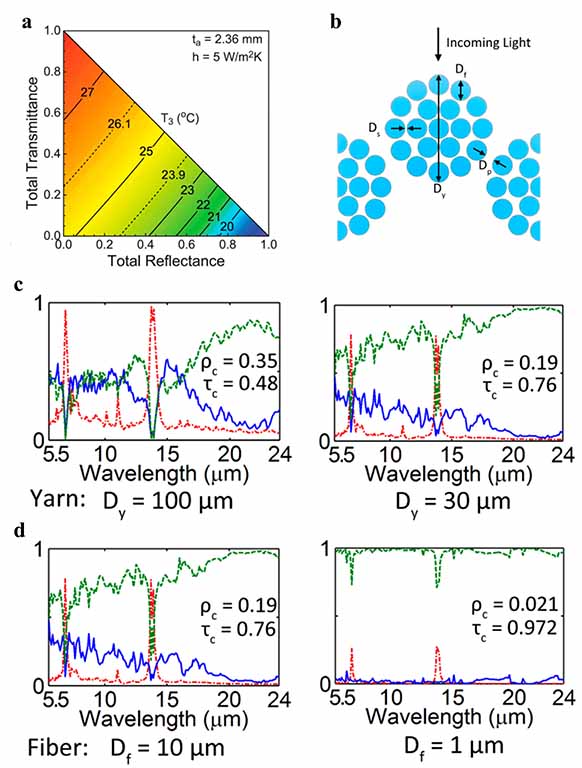}
	\caption{Cooling textile designed by  infrared-transparent visible-opaque fabric (ITVOF).(a) Calculated contour maps of the
maximum ambient temperature without sacrificing thermal comfort as a function of the total IR transmittance and reflectance of
textile in the conditions of the air convective heat transfer coefficient (h=3W/m$^{2}$K) and gap thickness (t$_{a}$=2.36 mm). (b) A
numerical finite-element simulation model was used to study the effect of polyethylene fabric structure on the IR optical properties.
In these simulations, these yarns were represented as parallel circular arrays. Incident light was set to be at normal incidence.
Specific input parameters D$_{y}$, D$_{f}$, D$_{s}$, and D$_{p}$ represent the yarn diameter, the fiber diameter, the fiber
separation distance and the yarn separation distance, respectively. (c) Numerically simulates the effect of reducing the size of the
yarn on the MIR optical properties of fabric at a fixed fiber diameter (D$_{f}$=10 $\mu m$). (d) Assess the impact of decreasing the
size of the fiber on the IR optical properties of fabric at a fixed yarn diameter (D$_{y}$=30 $\mu m$). The red, blue, and green
curves show the simulated absorptance, reflectance, and transmittance for the fabric with specific structure. \cite{ref24}}
\label{Fig.4*}
\end{figure}

\section{\label{sec3}Cooling textiles by regulating thermal radiation and conduction}In recent years, the heat stress has brought
health threat to the majority of people. To alleviate the thermal discomfort
caused by heat waves, the refrigeration equipments are used to provide indoor cooling, which consumes a large amount of
energy around the world annually. However, these equipments are low-efficiency temperature-regulating technology due to the waste of numerous resources in some inanimate objects. At present, the cooling textile has been regarded as a
promising technology because it emits the excess body heat to surroundings based on regulating the innate ability of the body. As a result, many cooling textiles have been proposed to overcome the adversity. In the following, few different methods of developing the cooling
textile through enhancing the MIR transmittance, NIR reflectance and thermal conductance of textiles will be summarized (Fig. \ref{Fig.3*}a).
\begin{figure*}
\centering
\includegraphics[scale=0.85]{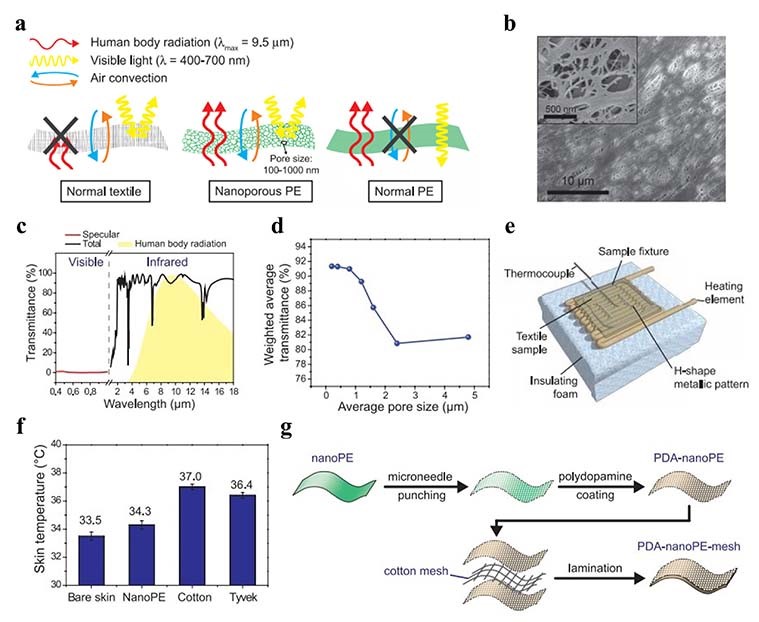}
\caption{Cooling textile with nanoporous polyethylene (nanoPE) film. (a) Schematics depict the comparison of the body heat
dissipation and air permeability of skin covered with normal textile, nanoPE, and normal PE, respectively. (b) The SEM
image of nanoPE film shows its small hole diameter that falls within the range of 50-1000 nm. (c) Simulated transmittance spectra of
nanoPE in the visible and infrared wavelengths. (d) Simulated weighted average transmittance of the nanoPE with different pore sizes
based on human body
radiation. (e) Schematic of the experimental measurement set-up. (f) Measured the simulational skin temperature for bare skin and
skin covered with nanoPE, cotton, and Tyvek respectively. (g) Schematic illustration of the fabrication process of PDA-nanoPE-mesh.
\cite{ref40}}
\label{Fig.5*}
\end{figure*}
\subsection{\label{sec3.1}Enhancing the MIR transmittance}According to the one-dimensional
steady-state-human heat transfer model mentioned in the Section \ref{sec2.3}, textiles can block the direct radiation heat transfer
of the body. Traditional textiles, such as cotton, wool and polyester, are invisible to electromagnetic waves in the MIR band. As
the chemical bonds of C-O, C-N, S-O and C-H contained in the chemical structure of traditional textiles exist in some vibration
modes, it produces several absorption peaks in the same area as the peak of human thermal radiation causing the strong absorption of
human body radiation \cite{ref14}. It is easy to accumulate heat around the body causing thermal discomfort in a hot environment. To
relieve the heat stress of the body in high temperature conditions, researchers devote to designing a novel type
of textiles with high thermal radiation transparency \cite{ref24}. In fact, the development of MIR transparent textile will be the best choices of the perfect emitter of human body ($\epsilon=0.98$) \cite{ref38},
which obtains the best heat dissipation effect.

For this purpose, Tong et al introduced the conceptual framework of the infrared-transparent visible-opaque fabric (ITVOF) that
utilizes the heat dissipation of thermal radiation emitted via the body into the environment directly to achieve cooling
\cite{ref24}. In this work, the heat transfer model (Section \ref{sec2.3}) was used to explore the requirement of IR optical properties of
the ITVOF with the better cooling effect. The ITVOF presented a minimum infrared transmittance of 0.644 and a
maximum infrared reflectance of 0.2 extending the range of thermal comfort from 23.9 to 26.1 $^\circ$C\ . Calculated results also
indicate that personal cooling textile can be achieved by maximizing MIR transmittance and minimizing MIR reflectance even under the
more insulating textiles with a higher air gap (t$_{a}$=2.36mm), as shown in Fig. \ref{Fig.4*}a. Additionally,
optimizing the diameters of both the fiber and yarn is also regarded as an effective method for enhancing the infrared transmittance of
fabric (Fig. \ref{Fig.4*}b-d).

As above metioned in the cooling textile, Cui's group has made a considerable progress experimentally. In 2016, they
designed a nanoPE film with high infrared transmittance, but it was opaque to visible light (Fig. \ref{Fig.5*}a) to enhance human thermal
comfort in hot conditions \cite{ref40}. The electromagnetic wave radiated by the body can efficiently radiate heat into the
environment through textiles. Moreover, nanoPE film is invisible to the human eye due to the fact that the size of the holes (50-1000
nm) embedded in the nanoPE is exactly the same as the wavelength of visible light (400-700 nm), resulting in Mie scattering (Fig.
\ref{Fig.5*}b). Although the size of these holes is comparable to the wavelength of visible light, its size (7000-15000 nm) is much smaller than the
MIR wavelength. Therefore, the nanoPE film still has high infrared transmittance and little impact on human
body heat dissipation (Fig. \ref{Fig.5*}c). The simulation results using rigorous coupled-wave analysis (RCWA) shows that the
transmittance of the material is more than 90\% in the wave band larger than 2 $\mu m$ (Fig. \ref{Fig.5*}d). In visible wavelengths,
however, there is a very low specular transmittance (the lower the specular transmittance, the higher the opaque) \cite{ref41}. In a
thermal experiment, nanoPE film, traditional cotton textiles and excellent non-woven fabrics Tyvek increase skin surface
temperature by 0.8, 3.5 and 2.9 $^\circ$C\ , respectively, which indicates that nanoPE film does possess excellent heat
dissipation performance (Fig. \ref{Fig.5*}e-f).

Although the nanoPE film presents a good heat dissipation performance, there still
exists some close-fitting wear properties that are not suitable for human body. For example, nanoPE film is pretty electronegative
while our body has a strong electronegativity, which easily causes electric contact under the condition of the skin friction with
clothings \cite{ref42}. Peng et al. improved this design by adopting a mass-produced fiber extrusion technology that firstly
squeezed the nanoPE into fibers obtaining the knitted/woven fabrics \cite{ref43}. The nanoporous polyethylene woven fabric (nanoPE
fabric) greatly improves the wearing comfort compared with the nanoPE film, which is attributed to the knitted/woven structure.
Fortunately, the nanoPE fabric still maintains an excellent optical property, further accelerating the commercialization of textiles
based on the preparation of nanoPE materials.
\begin{figure*}
\centering
\includegraphics[scale=0.48]{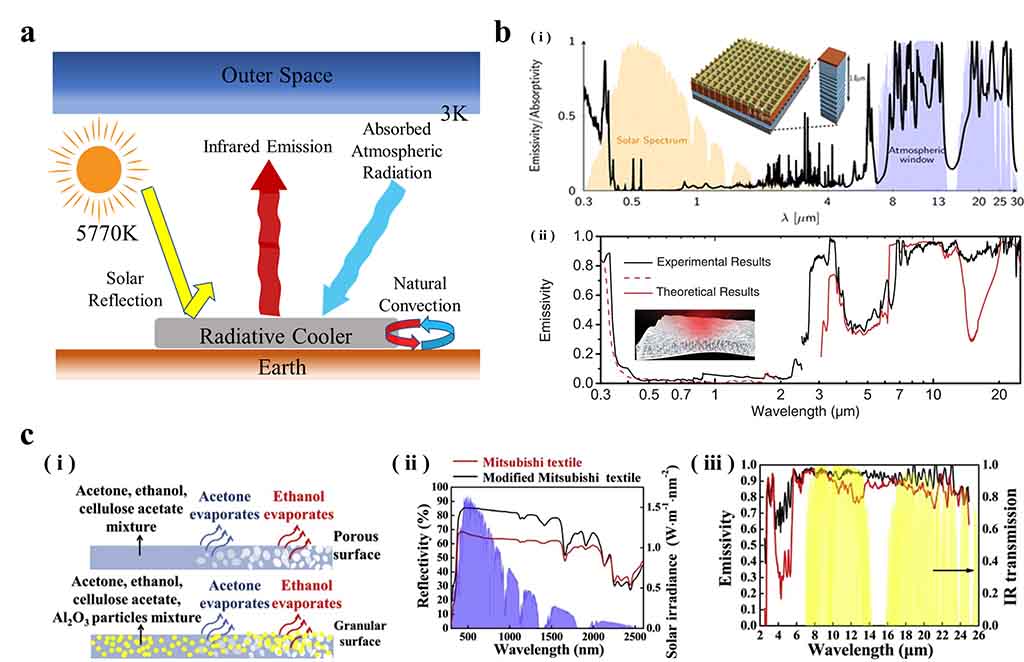}
\caption{The novel idea of cooling textiles inspired by the concept of radiative cooling. (a) Principles of radiative cooling. (b)
Different design ideas of the daytime radiative cooling device. (i) Radiative cooler with multi-layered periodic hole array
micro-structure \cite{ref37}. (ii) The radiative cooling film is achieved by embedding SiO$_{2}$ microspheres in the polymer film
\cite{ref50}. (c) The Al$_{2}$O$_{3}$--cellulose acetate coated textile for accelerating the heat dissipation of the human body
\cite{ref61}. (i) Schematic illustration of the phase inversion process of  the coating of cooling textile. Measured the NIR reflectance
(ii) and MIR emittance (iii) of cooling textile.}
\label{Fig.6*}
\end{figure*}

Infrared-transparent textiles have held great promise for regulating the heat dissipation of human body, but the colored cooling textiles without compromising the infrared properties remain the major challenge in the commercialization process. This is due to the fact that the chemical bonds of traditional organic dye molecules intensely absorb human body radiation. Fortunately, cai et al. proposed firstly an infrared-transparent textile whose coloration can be adjusted by controlling the diameters of inorganic nanoparticles based on the Mie resonance responses \cite{ref25}. The colored textile demonstrated the high MIR transmittance of 0.8 and the excellent cooling effect of 1.6-1.8  $^\circ$C\ , which provides an important step toward the practical applications of advanced textiles.

\subsection{\label{sec3.2}Enhancing the MIR emittance}The objects can emit heat via thermal radiation and those with higher temperature have a stronger emissive power. Consequently, the objects at the Earth's surface can transfer heat to the
outer space through the transparency window due to the large temperature difference between the object (above 300 K) and outer space
(3 K). The surface temperature of objects can be cooled that is the passive radiative cooling effect. In fact, radiative cooling is
a general phenomenon in our life. Such as the formation of dew or frost is a typical representation of radiative cooling progress.
Notably, the phenomenon can achieve the cooling effect without inputting any external energy. Thus, numerous researchers devoted many efforts into the applied research of radiative cooling. As shown in Fig. \ref{Fig.6*}a, the MIR emittance of cooler need to be
improved as much as possible to maximize the cooling effect of cooler, which enables much radiative heat to release into the outer
space through the transparency window. Importantly, the obtained energy of cooler from other pathways should be minimized. One of the
typical external energy inputs is solar energy, causing the temperature rise of cooler surface. Therefore, the cooler should also enhance the NIR reflectance to reduce the input of solar energy.

Nighttime radiative cooling devices have been widely studied decades ago
\cite{ref3,ref4,ref5,ref6,ref8,ref9,ref21,ref22,ref29,ref30,ref31,ref32,ref33,ref34,ref35}. These nighttime radiative coolers usually
consist primarily of some natural or synthetic materials, like paints \cite{ref3,ref4,ref5,ref6,ref8,ref9,ref21,ref22}, polymer films
\cite{ref29,ref30}, SiO films \cite{ref31,ref32,ref33,ref34,ref35} and other composite materials \cite{ref35,ref36}. Although these
devices demonstrate a relatively high MIR emittance in the transparency window, they lack the ability of reflected sunlight that can
not achieve an obvious cooling effect in daylight. Meanwhile,  the real cooling requirement is usually generated during the daytime. To achieve daytime radiative cooling, we need to develop the spectrally selective materials with high MIR emittance and NIR
reflectance. However, the unique materials have not been discovered in nature yet. With the development of the thermal radiation
theory of nanomaterials and nanostructures into free space, daytime radiative coolers have made an important progress in recent
years \cite{ref233,ref259}. In 2013, Fan and coworkers  theoretically designed a photonic crystal with the spectrally selective property,
which achieved firstly the concept of daytime radiative cooling \cite{ref37}. The photonic crystal is composed of multi-layered
micro-structure with periodic hole arrays obtaining the spectrally selective property. In consideration of the complex preparation of
surface microstructures, Fan et al. further developed the simplified photonic crystal composed of silver film, alternating HfO$_{2}$
and SiO$_{2}$ nano-layers in 2014 (Fig. \ref{Fig.6*}b(i)) \cite{ref49} . Above mentioned photonic crytal enables the better spectrally
selective ability and also achieves the remarkable cooling effect. But these coolers are hard to realize large-scale applications in
real life due to the processing difficulty and cost issues. To solve this issue, Yin et al proposed a  scalable-manufactured cooling
film by embedding SiO$_{2}$ microspheres in a polymeric matrix in 2017 (Fig. \ref{Fig.6*}b(ii)) \cite{ref50}. Zhu et al. fabricated a hierarchically nanofibre-based film with the selective properties that achieving the high relectance in the solar spectrum by controling the diameters of the nanofibres, which also provide a novel method for enabling the large-scale applications of all-day radiative cooler \cite{ref261}. Moreover, Li et al
demonstrated a cooling wood with a high practical value using the unique spectral features of the cellulose nanofibers in 2019 \cite{ref60}.

Intriguingly, as shown in Fig. \ref{Fig.1*}b, most of the human thermal radiation wavelengths fall within the atmospheric transmission
window. Hence, the concept of radiative cooling has the potential of applying to design the cooling textile, according to the heat transfer mechanism of cooling textile (Fig. \ref{Fig.3*}a) . Liu et al. made use of the above concept to design the cooling textile by coating Al$_{2}$O$_{3}$ dispersed-cellulose acetate with cotton (Fig. \ref{Fig.6*}c) \cite{ref61}. It is important for cellulose acetate to accelerate the radiative heat transfer with the high MIR emittance. In order to explore  more novel cooling textiles through enhancing the MIR emittance based on the radiative cooling principle, we still shoulder heavy responsibilities.

\subsection{\label{sec3.3}Enhancing the NIR reflectance}The aforementioned cooling textiles that improves the body's
radiation heat dissipation is beneficial for indoor sedentary workers to relieve thermal comfort caused by the high temperature. However, outdoor activities are also an
indispensable part in our daily life. For the heat source of the body, not only is it from its own metabolic heat, but also it
absorbs the heat from the sun when human do outdoor activities (Fig.\ref{Fig.1*} a). In summer, people who engage in outdoor activities are more likely to
suffer from heat stress \cite{ref47,ref48}. In contrast with indoor sedentary workers, the absorption of solar heat energy has a
significant influence on human thermal comfort.
\begin{figure*}
\centering
\includegraphics[scale=0.475]{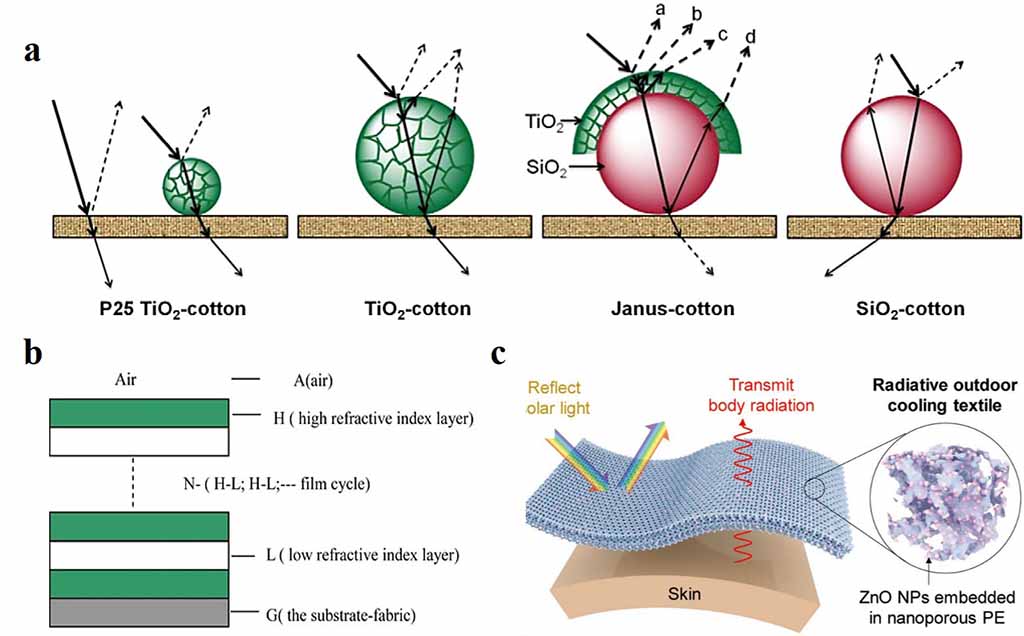}
\caption{Outdoor cooling textiles with different NIR reflective materials. (a) The different reflection mechanisms of P25
TiO$_{2}$-cotton, TiO$_{2}$-cotton, TiO$_{2}$-SiO$_{2}$-cotton and SiO$_{2}$-cotton, respectively \cite{ref174}. (b) A structure
diagram of the multi-layer film with high NIR reflectance. The system with different reflective index is coated with the cotton
alternatively \cite{ref170}. (c) An outdoor cooling textile is obtained by embedding the inorganic nanoparticles (ZnO) into the
nanoPE \cite{ref19}.}
\label{Fig.7*}
\end{figure*}

Solar heat energy is mainly generated in the NIR (700-1100 nm) \cite{ref155}. Most of the solar energy absorbed by
human body results in the higher body temperature, causing the decrease of human thermal comfort. Thus, minimizing the absorption of
solar energy to the human body plays a significant role in improving human thermal comfort (Fig. \ref{Fig.3*}a). Recently, a promising method applying
the NIR reflective materials in textiles has been proposed to achieve outdoor cooling textiles. Specifically, NIR reflective
materials can be classified as inorganic materials, organic materials and metallic materials, etc \cite{ref156}. Among the large
number of NIR reflective materials, inorganic materials have been extensively studied because they exhibit excellent weatherability,
heat stability and chemical inertness \cite{ref157}. One of the inorganic materials is rutile titanium dioxide that has a high NIR
solar reflectance of about 87.0\% due to its high refractive index. As a consequence, researchers proposed a novel cooling textile by
coating the TiO$_{2}$ nanoparticles with the cotton fabric \cite{ref159}.

Notably, both the particle size and refractive index are two major factors affecting the NIR reflectance of TiO$_{2}$ particles
\cite{ref155,ref162}. According to Fresnel's rules, TiO$_{2}$ particles are sized at 1/3 to 1/2 of the incident radiation wavelength
(350-550 nm) that is to be reflected, which obtains the maximum NIR reflect efficiency \cite{ref157,ref164,ref165}. Meanwhile, the NIR
reflectance of TiO$_{2}$ particles is also related to the value of refractive index. TiO$_{2}$ is mainly composed of rutile and
anatase with the refractive index of 2.73 and 2.55, respectively \cite{ref166}. Thus, it is possible to increase the reflection
efficiency of TiO$_{2}$ based on the phase transformation to rutile. Recently, Yau et al studied the UV-vis-NIR reflectance of
TiO$_{2}$ particles with different sizes and crystalline phases \cite{ref156}. These results stated that the refractive index and
particle diameter have a combined effect on the NIR reflectance of the TiO$_{2}$ powder. The highest solar reflectance occurred in
the TiO$_{2}$ sample with an anatase:rutile ratio of 35:65 and a particle diameter of 563 nm.
\begin{figure*}[htb]
\centering
\includegraphics[scale=0.5]{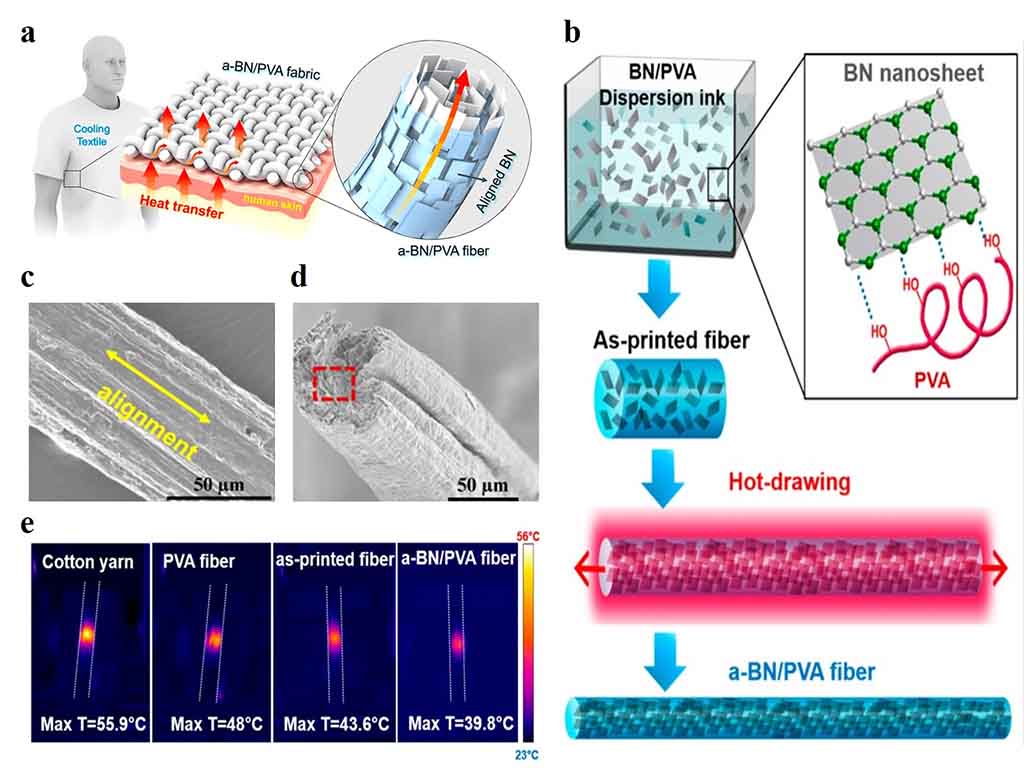}
\caption{Cooling textile is composed of aligned highly boron nitride/poly vinyl alcohol (a-BN/PVA) composite fibers fabricated by the
simple three-dimensional printing technology. (a) Schematic illustration of the cooling textile consisting of thermally conductive
a-BN/PVA fibers. (b) A schematic diagram of the specific fabrication process of a-BN/PVA composite fiber. (c), (d) SEM images of the
a-BN/PVA fiber. (e) Exploring the heat conduction characters of cotton, PVA, as-printed BN/PVA without BNNS alignment and a-BN/PVA
fibers with the same diameter by a laser-IR camera system. \cite{ref107}.}
\label{Fig.8*}
\end{figure*}

 In addition to the above methods, the multilayers consisting of  high and low refractive index materials have been proposed to improve the NIR
reflectance\cite{ref168,ref169,ref170,ref171}. Mennig et al. prepared multilayer NIR reflective coatings with the reflectance of
almost 72\% via coating six quarter wave thick layers using SiO$_{2}$ and TiO$_{2}$ nanoparticles of 10 and 4 nm diameters,
respectively  \cite{ref169}. In addition, Liu et al. further tested the reflection spectra of the multilayer TiO$_{2}$/SiO$_{2}$
film, indicating that the reflectance of the fabric with multilayer film was higher than that of the monolayer film \cite{ref170}
(Fig. \ref{Fig.7*}b). Moreover, they also indicated that the size of the component particles and the thickness of the deposited
layers have an essential effect on the NIR reflectance. Agrawal et al. explored the TiO$_{2}$-SiO$_{2}$ Janus particle combining
with the above two methods for enhancing the NIR reflectance  (Fig. \ref{Fig.7*}a) \cite{ref174}. Compared with TiO$_{2}$ particles
treated cotton fabric, TiO$_{2}$-SiO$_{2}$ Janus particles treated cotton fabric shows remarkable high NIR reflectance(79\% at 1000
nm). The Janus particles exhibits significantly higher NIR reflectance because it has two different refractive index materials, which
may be similar to multilayered coatings. The unique structure makes the reflection take place from multiple boundaries.
\begin{figure*}
\centering
\includegraphics[scale=0.5]{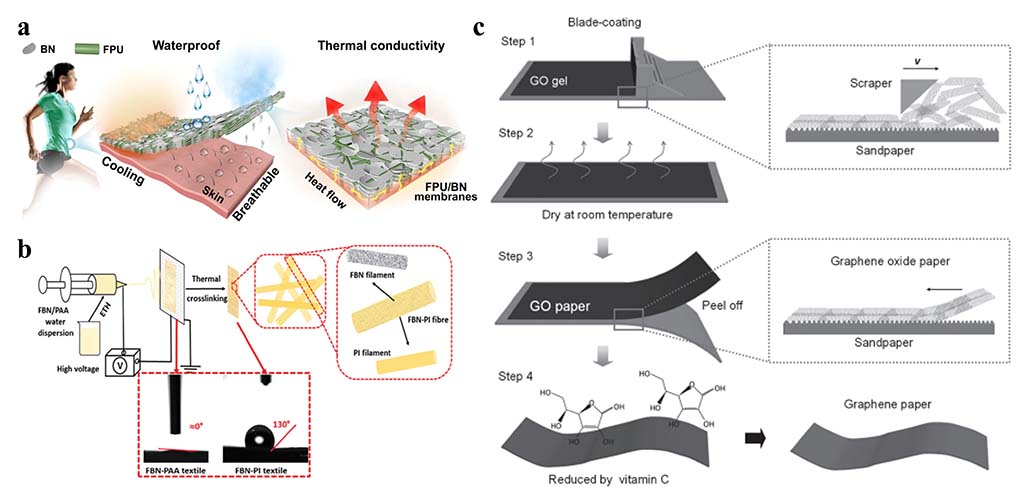}
\caption{Cooling textiles with different excellent heat conductance materials. (a) Schematic of the mutifunction textile comprised of fluorinated polyurethane (FPU)-BN nanosheets with thermoconductive, moisture-permeable, and superhydrophobic \cite{ref16}. (b) The cooling textile is composed of functional
boron nitride (FBN)-polyimide (PI) fiber fabricated by the electrospinning technology \cite{ref187}. (c) The preparation process of the graphene papers with high thermal conductance \cite{ref188}.}
\label{Fig.9*}
\end{figure*}

To prevent the human from the outdoor heat stress, we not only reduce the absorption of solar heat energy, but also improve
the radiative heat dissipation. This requires textile with a good spectral selectivity in the visible and mid-infrared bands.
However, the shortage of materials with strict spectral selectivity presents a challenge for the development of outdoor cooling
textiles. Cai et al. demonstrated for the first time a nanocomposite textile that reflects 90\% of sunlight and simultaneously
transmits a large amount of human thermal radiation \cite{ref19}. They proposed to embed zinc oxide nanoparticles into nanoPE
(ZnO-PE) to obtain a novel composite material with selective spectral response by combining the intrinsic material properties and
photonic engineering techniques. (Fig. \ref{Fig.7*}c). In typical outdoor environments (peak solar irradiance over 900 $W/m^{2}$), the
composite material could reduce the simulated skin surface temperature by about 10 $^\circ$C\ compared with traditional cotton
fabrics. Although the nanoPE has a high infrared transmittance, its relatively weak ability to reflect sunlight (refractive index is
about 1.5), which is insufficient to meet the outdoor human thermal comfort \cite{ref51}. Typical inorganic solid nano-ZnO particles
improved the refractive index of textile(n=2)  \cite{ref52}, without sacrificing the radiation heat dissipation due to their poor
electromagnetic wave absorption from visible to MIR band (400 nm-16 $\mu m$) \cite{ref53, ref54}. The combination of ZnO
nanoparticles and nanoPE is exactly what satisfies the requirements of strict spectral selectivity.

\subsection{\label{sec3.4}Enhancing the thermal conductance}As mentioned above, most researchers are paying attention closely to
design textiles that regulate body radiation to enhance human thermal comfort. However, there are few studies on adjusting the
thermal conductance of textiles to regulate the human body heat transfer. As we know, for the three main forms of heat transfer, heat
exchange with the outside environment through heat conduction is as important as the other two methods. Especially, heat
conduction plays a dominant role in the heat exchange systems between the skin and clothes. On the basis of Eq. (\ref{E9}), the thermal conductance of textile ($G_{f}$) exerts an indispensable influence on the inner and outer surface temperature of textile. Thus, the development of textiles
with high thermal conductance is also promising for relieving human heat stress in indoor environments. Nevertheless,
traditional textiles often have low thermal conductance, which seriously hinders the release of excess heat around the body into the
environment. For example, common textiles such as cotton, wool and nylon have lower thermal conductivity (0.07, 0.05 and 0.25 W/mK,
respectively) \cite{ref99,ref100}. Most of our daily textiles are porous, and air will be trapped in the pores of the clothes, which
will decrease the overall thermal conductance of the textiles. The easiest way to solve this problem is to block these pores with
the dense material, but these will greatly reduce the air permeability of the garment.

Hence, many researches have done on improving the thermal performance of cotton textiles, most of which focus on
adjusting the thickness, porosity, air permeability, and perspiration of cotton textiles \cite{ref102,ref103,ref104}. Improving the
heat dissipation of cotton fabrics using high thermal conductance materials has been reported in recent years. In 2017, Hu et al.
prepared a novel fiber by using  a fast and scalable three-dimensional (3D) printing method based on highly aligned boron nitride
(BN)/polyvinyl alcohol (PVA) with high thermal conductance (Fig. \ref{Fig.8*}a) \cite{ref107}. BN has been considered as an effective
material in thermal regulation applications due to its excellent thermal conductance yet electrical insulation. The two dimensional
BN nanosheets (BNNSs) have a planar thermal conductivity of up to 2000 W/mK, indicating that they have a better thermal management ability of in the
system \cite{ref108,ref109}. In this design, the composite fiber with highly oriented BNNSs provided a large number of heat transfer
channels for the body heat dissipation (Fig. \ref{Fig.8*}b-d), effectively improving the heat dissipation of the textile. A
laser-infrared camera system is used to investigate the heat transfer performance of different fiber samples with the same diameter.
Experimental results show that a-BN /PVA composite fiber does have excellent thermal conductance, which can transfer heat quickly
and reduce its temperature (Fig. \ref{Fig.8*}e).

Up to now, researchers have made some new progress using the ideal heat conductance of BN. Such as Fu and coworkers
added high content of edge-selective BNNSs to a biodegradable cellulose/alkaline/urea aqueous solution achieving a high axial thermal
conductivity of 2.9 W/mK \cite{ref185}. Strikingly, the BNNS also were enriched in the hydrophobic fluorinated polyurethane to construct the moisture-permeable, superhydrophobic, and cooling fabrics using the simple electrospinning technique (Fig. \ref{Fig.9*}a) \cite{ref16} . Although these textiles have achieved excellent thermal conductivities based on BNNS, these materials can not limit high temperatures \cite{ref107,ref186}, which will bear their further applications
in an extremely high temperature environment. As a result, Wang et al proposed firstly a high temperature thermally conductive
textile based on functional BN nanosheets and PI nanocomposite fibers. They found that the nanocomposite textile revealed a high thermal
conductivity of 13.1 W/mK at 300 $^\circ$C\ (Fig. \ref{Fig.9*}b) \cite{ref187}. Besides, carbon nanotubes have been widely used in thermal management systems due to their unique thermal properties. For example, typical thermal conductivities of single-walled and multi-walled carbon nanotubes  are measured as high as 3000-3500 W/mK and 500-2800 W/mK at room temperature \cite{ref105}. Thus, Abbas et al. coated carbon nanotubes into cotton textiles to improve the thermal conductance
of cotton fabrics \cite{ref101}. The thermal conductance of the coated cotton increases with the increase of the carbon nanotube content in the coating. The thermal conductance of textile coated with 11\% multi-walled carbon nanotube coating on cotton fabric can be increased by 78\% compared with
untreated cotton. Moreover, Guo also designed a textile with personal thermal regulation based on the high out-of-plane thermal
conductance of the graphene paper (Fig. \ref{Fig.9*}c) \cite{ref188}.

At the same time, by strengthening thermal convection and sweat evaporation can also help improve the thermal conductance of textile. Specifically, strengthening thermal convection is equivalent to improving the effective thermal conductance between the skin and the environment to cool down the human body, which is shown in Section \ref{sec2.3} in details. For example, Zhao et  al. designed the ventilated garment consisted of a short sleeve jacket and two ventilation units, and each ventilation unit was a small fan in experiment \cite{ref251}. Meanwhile, some works strengthed the cooling mechanism of sweat evaporation  by designing the wicking performence of textiles \cite{ref253,ref254,ref255}. Intriguingly, Cui et  al. proposed a novel integrated cooling textiles combining thermal conduction with sweat evaporation \cite{ref256}. The ingenious design accelerated the heat transfer rate that satisfies the thermal comfort of human in high humidity circumstance by improving the effective thermal conductance of textile.

\section{\label{sec4}Warming textiles by modulating thermal radiation and conduction}Compared with the cooling textiles used in hot weather, warming textiles maintain comfortable body temperature in cold conditions. Indeed, we have to heat
indoor space to maintain human thermal comfort using air conditioner and heater in the cold winter, which will
inevitably consume much energy. Further, studies have shown that the annual energy consumption for indoor heating accounts
for 22.5\% of building energy and it has exceeded the energy for indoor cooling (14.8\%)  \cite{ref55}. To reduce the energy
consumption of indoor heating, scientists payed closely attention to study functional materials with high reflectance and low emittance \cite{ref175,ref176,ref177,ref79}. However, there are still a large portion of resources wasted to maintain the temperature of
empty space and inanimate objects. According to the Fig. \ref{Fig.3*}b, warming textiles effectively overcome these limitations as they can regulate the personal microclimate of the human body.

\subsection{\label{sec4.1}Increasing the MIR reflectance}At present, there are several warming textiles using metals with high MIR reflectance in the field of textiles. Namely, we can improve the MIR reflectance ($\rho_{i}$) of inner surface of textile, as shown in Eq. (\ref{E4}) and (\ref{E5}). However,  these warming textiles have some shortcomings. For instance, the space
blanket consists of a plastic sheet (e.g., polyethylene terephthalate) overlaid with a thin continuous layer of metal (e.g, aluminum)
that lacks breathability, making it uncomfortable for daily wear \cite{ref56}. The Omni-Heat technology, a commercial technology in
recent years, which prints the metallic dots onto the inside of garments to reflect human thermal radiation, has low reflectance
actually \cite{ref57}. Therefore, researchers devote themself to developing some warming textiles with high warming performance
and wearability.
\begin{figure}
\centering
\includegraphics[scale=0.39]{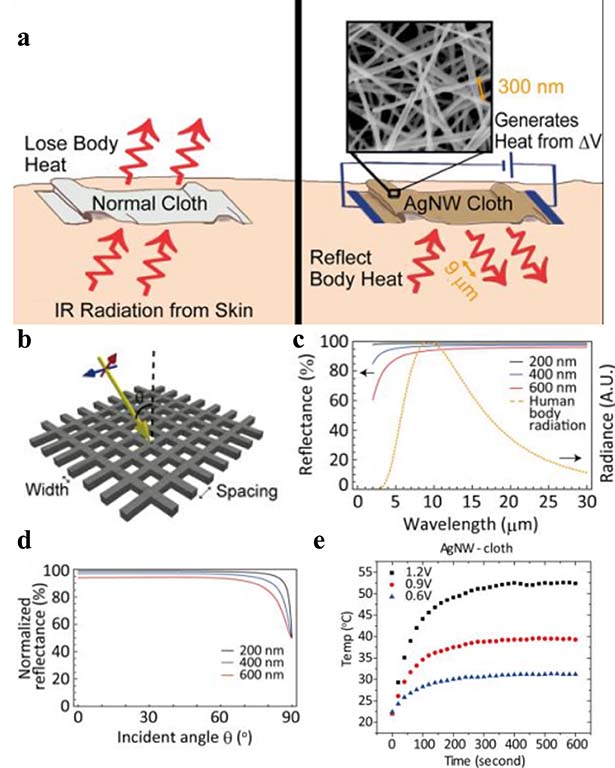}
\caption{Warming textile with metallic silver nanowires (AgNW) . (a) Schematics illustrate the comparison of the heat transfer
mechanism between normal cloth and AgNW-cloth. The AgNW-cloth provides extra heat for body through applying with voltage. (b) A
diagram of the simulational model used to describe the IR reflectance of a metallic AgNW network. (c) The comparison of calculated IR
reflectance spectral  of AgNW network with the space of 200, 400, and 600 nm at an abnormal incidence, respectively. Yellow dotted
line represents the human body IR radiation spectrum. (d) Caculated IR normalized reflectance spectra of AgNW network with the space
of 200, 400 and 600 nm based on the body thermal radiation from normal incidence to horizontal incidence. (e) The temperature change
of AgNW-cloth versus time after inputing different voltage. \cite{ref58}}
\label{Fig.10*}
\end{figure}
\begin{figure*}
\centering
\includegraphics[scale=0.5]{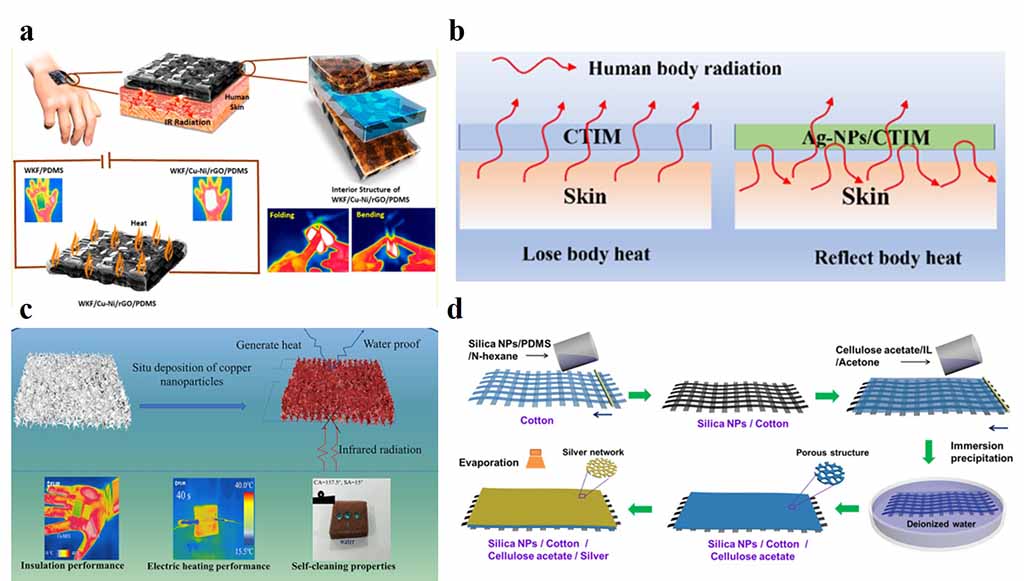}
\caption{Warming textiles with different high-reflectance materials. (a) WKF/Cu-Ni/rGO/PDMS warming textile \cite{ref179}. (b)
Ag-NPs/CTIM warming textile  \cite{ref180}. (c) Cu-MES warming textile \cite{ref181}. (d) Silica NPs/Cotton/Cellulose acetate/Silver
warming textile \cite{ref182}.}
\label{Fig.11*}
\end{figure*}

Hsu et al. demonstrated a novel warming textile to achieve this goal by coating with metallic silver nanowires on the cotton
\cite{ref58}. Fig. \ref{Fig.10*}a illustrates the effective mechanism of personal thermal management with the metallic silver nanowires
textile. The normal cloth provides little radiative insulation due to its relatively high emittance. However, novel textile after
coating the metallic silver nanowires forms a metallic conducting network reflecting effectively the majority of human body radiation. The spacing between nanowires networks is almost between 200 and 300 nm that is far less than the human body thermal
radiation wavelength. Thus, the textile with metallic nanowires seems to be a continuous metal film with high reflectance. Moreover,
the spacing of the nanowires network is much larger than the water molecule, which can enhance the water vapor permeability.
Specifically, researchers also proposed a simplified model to simulate the metallic silver nanowires network and studied its MIR optical
properties (Fig. \ref{Fig.10*}b). As shown in Fig. \ref{Fig.10*}c, the reflectances of the metallic silver nanowires networks with
different spacings are all high reflectance in the human body radiation spectrum, assuming normal incidence. Besides, the metallic
nanowires conducting network also achieved the active warming body by using Joule heating (Fig. \ref{Fig.10*}d-e). Notably, the
metallic silver nanowires textile was heating up to 38 $^\circ$C\ through only inputting 0.9 V voltage.

Although the silver nanowire-embedded textile successfully provides thermal insulating capability without sacrificing the
breathability, the commercialized applications of silver nanowires are difficult due to its extremely high cost. On the
contrary, copper is relatively cheaper and still has only 6\% lower conductance than silver. Hence, Hazarika et al. fabricated a
Woven Kevlar fiber (WKF) by directly growing vertical copper-nickel (Cu-Ni) nanowires (NWs)
on the WKF surface (Fig. \ref{Fig.11*}a) \cite{ref179}. In this design, Cu-Ni NWs with a tunable Ni layer on Cu core NWs greatly
enhance the oxidation resistance of the Cu NWs. In short, the usage of metal nanowires is a promising method for fabricating the
textiles with personal thermal regulation. Meanwhile, the preparative process of metal nanowires is pretty complicated, which adverses
to mass production.
\begin{figure*}
\centering
\includegraphics[scale=0.52]{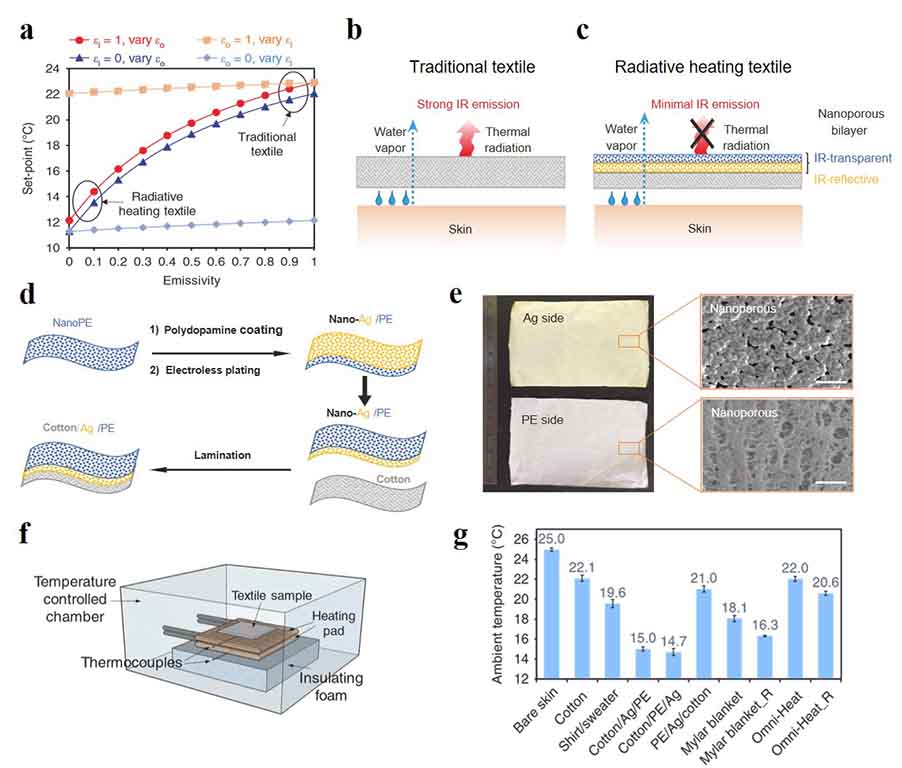}
\caption{Warming textile with nanoporous metalized polyethylene textile (nano-Ag/PE). (a) Calculated set-points of environmental
temperature for satisfying body thermal comfort versus the inner surface ($\epsilon_{i}$) and outer surface ($\epsilon_{o}$) IR
emittance of textile by using the heat transfer model. Schematics of the heat transfer of the human body covered with (b)
traditional textile and (c) nano-Ag/PE textile. (d) A diagram of the fabrication process of nano-Ag/PE textile. (e) Optical and SEM
images of the Ag side and PE side of the nano-Ag/PE textile show that it has better breathability. Scale bar, 1 $\mu m$. (f)
Schematic the experimental equipment. (g) The comparison of measured set-points of  environmental temperature for bare skin
and skin covered with different textile samples. Notably,  the left side of three systems with cotton/Ag/PE, cotton/PE/Ag, and
PE/Ag/cotton is in contact with the skin and the right side is facing the environment. Mylar blanket-R and Omni-R are the reversed
Mylar blanket and Omni blanket. \cite{ref23}}
\label{Fig.12*}
\end{figure*}

As a consequence, researchers demonstrated some warming textiles using simple technical operations. Yue et al. designed the Ag
nanoparticles infrared radiation reflection coating with high infrared reflectance by anchoring the sphere-like Ag nanoparticles on
the surface of waste paper cellulose fibers. The reflection coating also provides an excellent antibacterial activity due
to the existence of Ag-NPs (Fig. \ref{Fig.11*}b) \cite{ref180}. Similarly, Zhou et al. showed a multifunctional copper coated
melamine sponge (Cu-MES) with high infrared reflectance, excellent conductance, as well as high self-cleaning performance by
depositing the copper nanoparticles on the melamine sponge surfaces for personal thermal management applications (Fig.
\ref{Fig.11*}c) \cite{ref181}. Besides, Liu et al. fabricated a multifunctional cloth based on a normal cotton textile, which is designed by depositing a metallic
film on the nanoporous structure on one side and adding a hydrophobic coating (Fig. \ref{Fig.11*}d). The multifunctional cloth
guarantees the high MIR reflectance without sacrificing breathability and flexibility. More significantly, it also has other desired
properties including waterproofness, antibacterial activity,  and the ability to generate Joule heat \cite{ref182}.
\begin{figure*}[t]
\centering
\includegraphics[scale=0.47]{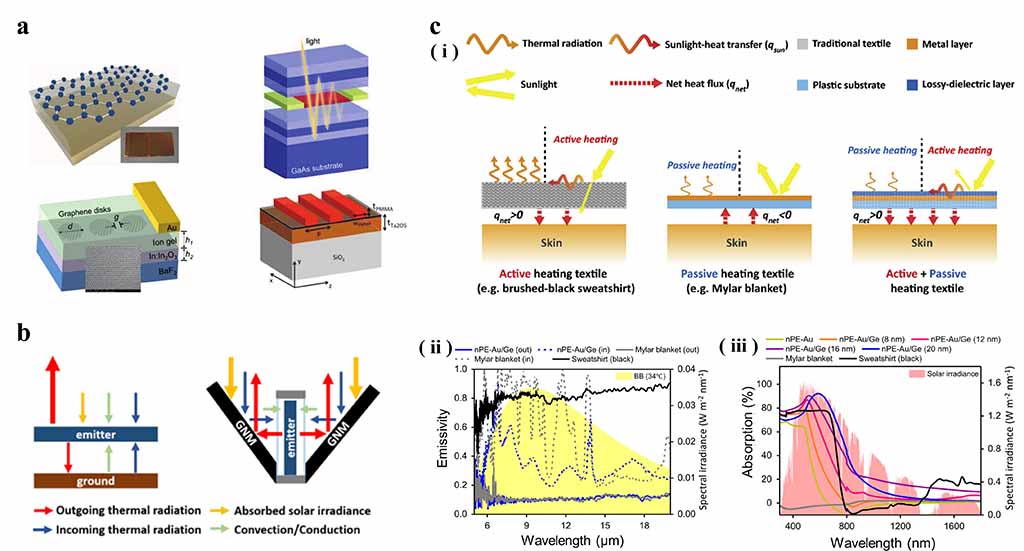}
\caption{The high NIR absorptance materials incorporated textiles. (a) Various designs to enhance the NIR absorption of materials
\cite{ref63,ref64,ref66,ref72}. (b) A double-sides radiative cooler with high NIR absorptance materials (graded nanocomposite
metamaterials) \cite{ref70}. (c) A colored outdoor warming textile is fabricated based on the concept of using the high NIR
absorptance materials. (i) The comparison of the warming mechanism of different textiles. Measured the MIR emittance (ii) and NIR
absorptance (iii) of the different textiles. \cite{ref196}}
\label{Fig.13*}
\end{figure*}
\subsection{\label{sec4.2}Decreasing the MIR emittance} Warming textiles achieved by increasing the MIR reflectance of
materials do not obtain a satisfactory insulating effect. Hence, in order to design textiles with thermal insulation properties, Cai et
al. found that textiles with low infrared emittance on the outer surface, which are more resistant to heat radiation from the human
body than those with high internal surface infrared reflectance based on the one-dimensional steady-state heat transfer model
\cite{ref23}. In this model, they studied respectively the influence of the inner ($\varepsilon_{i}$ ) and outer surfaces infrared emittance of the textile ($\varepsilon_{o}$ ) on the set-point of environmental temperature with maintaining the human thermal comfort, which are corresponding to Eq. (\ref{E6}) and (\ref{E9}). Meanwhile, Fig. \ref{Fig.12*}a shows that the set-point decreases monotonically with reducing $\varepsilon_{o}$ at fixed $\varepsilon_{i}$ ,
while it almost remains constant when $\varepsilon_{i}$ is changed at fixed $\varepsilon_{o}$.The
result is contradictory to the concept that can improve the effect of warming performance by reflecting back the infrared radiation
from the human body. In fact, the little impact of the inner surface infrared reflectance on warming is mainly attributed to the
dominancy of heat conduction over radiation in the heat transfer between the skin and the textile inner surface. Hence, the effect is
not significant by adjusting the infrared reflectance of the inner surface of textiles to improve the insulation capacity of
textiles. Nevertheless, reducing the infrared emittance of the textile outer surface can effectively suppress the human heat
radiation to the environment because heat radiation performs a dominant role in the heat transfer between the textile outer surface
and the environment.

In this regard, they applied a metal layer to the dopamine-treated nanoPE by electroless plating to form a composite film
\cite{ref59}. Then,  the composite film was laminated on cotton textiles, obtaining a novel textile with thermal insulation
effect (Fig. \ref{Fig.12*}d) \cite{ref23}. Coating the metal layer on a nanoPE substrate will naturally form a metal reflective layer
containing nanopores (Fig. \ref{Fig.12*}e). The size of the small holes in the metal layer is smaller than the infrared wavelength but
larger than the water molecules. Not only it has the effect of high reflection of infrared radiation, but it overcomes the
shortcoming of poor breathability of the space blanket (Fig. \ref{Fig.12*}b-c). What's more, another key point of the new textile is
that the nanoPE is coated with a metal layer, which exerts a simple protective effect on the metal layer. Meanwhile, the nanoPE can
also strongly scatter visible light due to the presence of nanopores in the material, improving the opaque of the textile.  Most
importantly, the nanoPE is placed on the outer layer that takes advantage of its low infrared radiation rate to reduce heat
dissipation. Indeed, experimental results indicated that the novel textile has a better thermal insulation effect (Fig.
\ref{Fig.12*}f-g).

\subsection{\label{sec4.3}Increasing the NIR absorptance} Although the warming textiles mentioned above have provided an effective
approach to maintain body temperature, they seem more appropriate for sedentary people in an indoor environment. As the same time, outdoor activities
are also of the utmost importance in humans even in cold winter days. The human body is more susceptible to catching colds as the
outdoor heat loss of the body is more severe compared with the indoor environment. It also poses great challenges for the development of outdoor warming textiles. Fortunately, outdoor solar irradiation has been extensively applied in our lives, such as solar
water heater, solar cell, and solar thermophotovoltaics, etc. These applications need to have a
solar selective absorber with high NIR absorptance. To make effectively use of solar energy, various methods have been proposed to
enhance the NIR absorptance of materials in recent years (Fig. \ref{Fig.13*}a). For instance, the studies concentrated on
enhancing the optical absorption in Graphene and some novel 2D materials  \cite{ref62}. Multiple designs have been proposed for improving
the NIR absorption of 2D materials, including metallic reflectors \cite{ref63,ref64}, plasmonic nanostructures \cite{ref65,ref66} and
photonic crystal nanocavities \cite{ref67,ref68,ref69,ref72,ref214}. Additionally, Zhou et al. designed a double-sided radiative cooler that
integrated the cooling and heating in a device using the graded nanocomposite metamaterials to absorb solar energy (Fig. \ref{Fig.13*}b) \cite{ref70} . Meanwhile, Chen et al. realized the stack of colored absorbers composed of TiN$_{x}$O$_{y}$ absorbing layer and
TiO$_{2}$/Si$_{3}$N$_{4}$/SiO$_{2}$ dielectric that can be applied in flexible substrates \cite{ref71}.
\begin{figure*}
\centering
\includegraphics[scale=0.55]{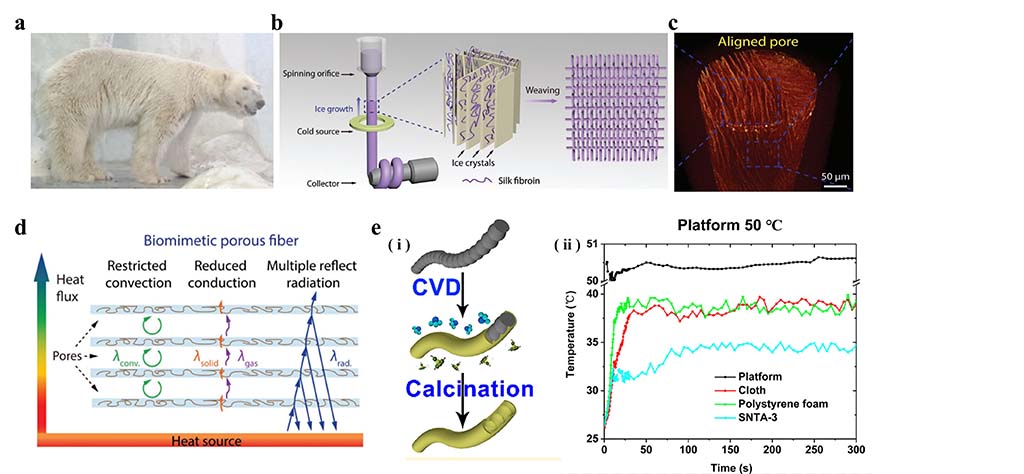}
\caption{Warming garments with low heat conductance materials. (a) Photo of the polar bear in the freezing conditions. (b) Schematic
depicts the fabrication process of the ``freeze-spinning'' technique of biomimetic fibers. (c) An X-ray computed microtomography
image of biomimetic fibers shows the aligned porous structure along its axial direction. (d) Schematic illustrates the heat transfer
mechanism of the biomimetic fibers with aligned porous structure \cite{ref189}. (e) Multifunctional nanotube aerogels is a potential
candidate material of warming textiles. (i) Schematic illustration of the fabrication process of multifunctional nanotube aerogels.
(ii) Measured the thermal equilibrium temperature of different materials \cite{ref77}.}
\label{Fig.14*}
\end{figure*}

Currently, Luo et al. reported a colored warming textile by coating a lossy dielectric layer and a metal layer with nanoporous
textile for indoor and outdoor environments simultaneously (Fig. \ref{Fig.13*}c) \cite{ref196}. The warming textile enabled a
remarkable warming effect of 6.3 $^\circ$C\ compared with the traditional textile by improving the NIR absorption and reducing the
MIR radiative loss of the human body simultaneously. As shown in Fig. \ref{Fig.13*}c, the metal layer of the structure greatly reduced
the outer surface emittance and the lossy dielectric layer, which leads to a large amount of the solar absorption. More importantly, its
color can be adjusted through changing the deposition time of the Ge coating layer due to the spectrally-selective broadband solar
absorption of the dielectric coating. As a result, the method that utilizes solar energy to enhance the warming effect of textiles
has the potential to promote the development of outdoor warming textiles (Fig. \ref{Fig.3*}b). Nevertheless, the relevant studies are still relatively
few based on the concept of improving the NIR absorptance of textiles. There is more space for us to explore high performance
outdoor warming textiles.

\subsection{\label{sec4.4}Decreasing the thermal conductance}In addition, researchers also designed the warming textiles with low thermal conductance. In other words, we can decrease the thermal conductance ($G_{f}$) of textile to achieve warming textiles, according to Eq. (\ref{E9}). In fact, air is the most common material with low heat conductance. As we all know,  polar
bears keep warm and survive using their specifically hollow hairs in an extremely cold environment (Fig. \ref{Fig.14*}a).
Consequently, fibers containing a porous structure could be a promising method to enhance the insulating properties of textiles.
Aerogel is regarded as a promising candidate for fabricating the fibers of warming textiles due to its ultra-low density, high
porosity, and large surface. In 2014, Si et al. developed a method for creating nanofibrous aerogels by combining the
electrospun nanofibres technology and fibrous freeze-shaping technology \cite{ref191}. In 2018, Du et al. fabricated the
multifunctional nanotube aerogels that can be used to provide light management and thermal insulation (Fig. \ref{Fig.14*}e) \cite{ref77}. The successful preparation of nanofibrous aerogels accelerated the development of warming textiles.

Up to now, many warming textiles based on the aerogel fibers have been proposed. Cui and coworkers presented a ``freeze-spinning''
technique by mimicking polar bear hairs, achieving continuous and large-scale fabrication of silk fibroin fibers with aligned porous
structure (Fig. \ref{Fig.14*}b-c) \cite{ref189}. The aerogel fibers show excellent thermal insulation property. Theoretically, the Ref (23) regard as the
thermal conductance of the aerogel fiber is the sum of thermal convection,
solid, and air thermal conduction, and thermal radiation.  (Fig. \ref{Fig.14*}d) \cite{ref190}. Specifically, air convection can be neglected because the air is blocked in those micropores. The thermal conductance
of the aerogel fiber with a porous structure will be reduced significantly due to the low heat conductance of air in comparison with
solid. In addition, the aerogel porous fiber with aligned porous structure can improve the reflectance of infrared light with its
numerous solid-air interfaces, which also inhibits the heat loss in the manner of thermal radiation. Furthermore, it is reported that
the aerogels with low thermal conductance are similar to that of air at ambient conditions, making the aerogels favorable for
effective thermal insulation \cite{ref191}.
\begin{figure*}
\centering
\includegraphics[scale=0.6]{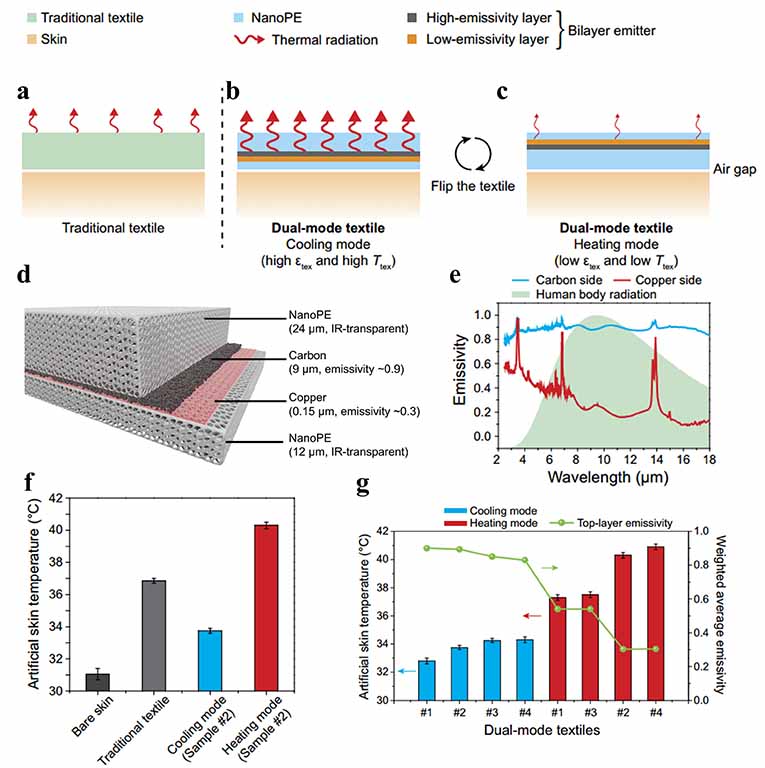}
\caption{The dual-function textile embeds a bilayer emitter in the nanoporous polyethylene (nanoPE) layer for human body warming and
cooling. Schematics depict the comparison of the heat transfer mechanism between (a) traditional textile, (b) dual-mode textile in
cooling mode, (c) and dual-mode textile in warming mode. (d) A diagram of the dual-mode textile structure. Notably, the nanoPE
thickness of upper and lower layers of the bilayer emitter is asymmetric which better controls the temperature of the emitter. (e)
Measured the emissivities of carbon and copper coating. (f) The comparison of the artificial skin temperature covered with various
samples: bare skin, traditional textile, cooling mode textile, and warming mode textile. (g) Measured the artificial skin temperature
covered with different four dual-mode textiles. Four dual-mode textiles are composed of the materials with four different emittance
combinations. The green line depicts the artificial skin temperature versus the top-layer emittance of dual-mode textile, which
presents an opposite trend. \cite{ref75}}
\label{Fig.15*}
\end{figure*}

Nowadays, more researchers have devoted themselves to developing aerogel fibers, such as graphene \cite{ref192},
silica \cite{ref193}, and cellulose aerogel fibers \cite{ref194}. However, these aerogel fibers either lack of approciate
mechanical properties or have a complex preparation process. The method of wrapping a protective layer with aerogel fibers has
been used to enhance the mechanical properties of aerogel fibers. In 2019, the work reported the cellulose acetate/polyacrylic acid
(CA/PAA)-wrapped silk fibroin (SF) aerogel fibers and cellulose nanofibrous (CNF)-wrapped cellulose-rich aerogel fibers fabricated by
the above method that present excellent mechanical and thermal insulating properties with a working temperature from -20 to 100
$^\circ$C\  \cite{ref204,ref205}. Although aforesaid aerogel fibers have basically satisfied our daily needs, the materials fail to
maintain the excellent thermal insulating properties at a temperature higher 200 $^\circ$C\ . To further expand
applications of aerogel fibers, Liu et al fabricated a nanofibrous Kevlar aerogel threads that were woven into textiles
illustrating the excellent thermal insulation and remarkable mechanical properties under extreme temperature (-196 or +300 $^\circ$C\
) and at room temperature \cite{ref195}. Fan et al. designed SiO$_{2}$/polyimide(PI) composite aerogel fibers that also present a low
thermal conductivity of 35 mW/mK at 300 $^\circ$C\ , which can be applied in many militaries and civil fields
\cite{ref206}.

\section{\label{sec5}Dual-mode textiles by adjusting thermal radiation and conduction}Cooling and warming textiles can be achieved
through modulating the transmittance, emittance, reflectance, and thermal conductance of textiles. In addition, the human body need the textile with cooling and warming characteristics to combat the changeable climate. Obviously, the design
mechanism behind the textiles is corresponding opposite, which brings a huge challenge. Hence, some researchers have done some explorations to achieve the dual-mode advanced textiles, as shown in Fig. \ref{Fig.3*}c.

\subsection{\label{sec5.1}Statically switching the MIR emittance}Importantly, Hsu et al. have achieved a dramatic breakthrough in
2017 \cite{ref75}. They embed a bilayer emitter inside a nanoPE designing a statically dual-mode textile (Fig. \ref{Fig.15*}d) This
dual-mode textile can easily switch modes between warming and cooling by flipping inside and outside. This work emphasized that the
regulation of human thermal radiation of textiles is not only related to the optical properties of the textile to the mid-infrared
electromagnetic waves but also connected to the temperature of the outer surface of the textile based on Stefan-Boltzman Law (Eq. (\ref{E6})).

In this dual-mode textile, Hsu et al. embedded double-layer emitter and nanoPE with different thickness can switch the outer surface emittance and temperature of the textile, respectively. In the
cooling mode, the porous carbon layer with an emittance of up to 0.9 (Fig. \ref{Fig.15*}e) faces the environment to increase the
surface emittance of the textile (Fig. \ref{Fig.15*}b). Moreover, the small thickness of nanoPE (12 $\mu m$) between the skin and the
emitter ensures a higher temperature on the textile surface. As a result, the combination of high emittance (high $\epsilon_{o}$)
facing outside and short emitter-to-skin distance (high $T_{o}$) will result in the high heat flow of the body through the textile.
On the contrary, the combination of metallic copper with a lower emittance of 0.3 facing outside (Fig. \ref{Fig.15*}e) and the
thicker nanoPE (24 $\mu m$) achieves a functional thermal insulation performance (Fig. \ref{Fig.15*}c). Experimental results demonstrate
that the double-mode textile indeed has a cooling effect of 3.1 $^\circ$C\ lower than that of the traditional textile. Meanwhile, it
can also obtain a warming effect of 3.4 $^\circ$C\ higher than that of the traditional textile (Fig. \ref{Fig.15*}f-g).
\begin{figure}
\centering
\includegraphics[scale=0.243]{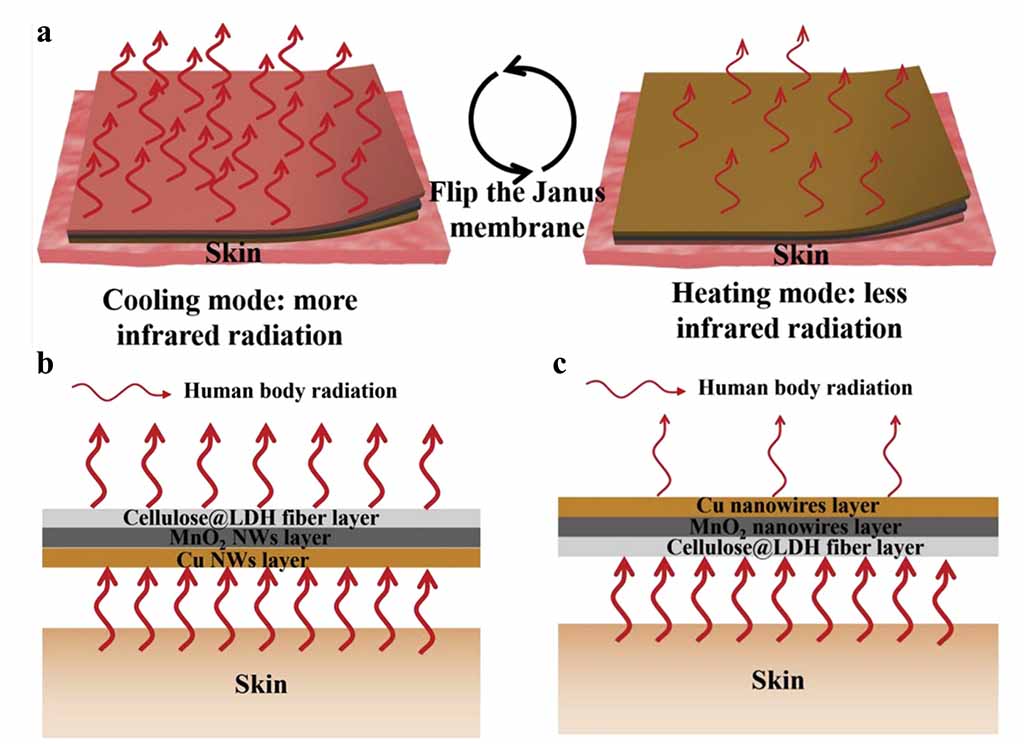}
\caption{The textile with warming and cooling abilities is comprised of a trilayer Janus membrane with a sandwich structure through
the technology of the vacuum filtration. (a) The Janus membrane provides the body with dual-mode by flipping the membrane. The
specifical schematics of the (b) cooling mode and (c) warming mode of the trilayer Janus membrane. \cite{ref183}}
\label{Fig.16*}
\end{figure}

According to the similar mechanism, Yue et al. developed a multifunctional Janus Cu/MnO$_{2}$ cellulose@Layered Double hydroxide fiber
(CMCFL) membrane with sandwich structure for on-demand personal thermal regulation (Fig. \ref{Fig.16*}a) \cite{ref183}. In this
structure, MnO$_{2}$ is placed into the middle layer to achieve the strong bonding and high interfacial compatibility of Janus
membrane. The main function of the Cu nanowires is an infrared reflector for warming. Additionally, the Cu nanowires layer also
provides the additional Joule heat and the antibacterial performance (Fig. \ref{Fig.16*}c). The cellulose@Layered Double hydroxide
fiber plays an essential role as a high infrared emittance layer (Fig. \ref{Fig.16*}b). Hence, the Janus membrane presents two
different infrared radiation properties, which further achieves the dual-mode function for the body. Besides, the sandwich structure
can also supply additional Joule heat using only a low voltage that combines passive warming with active warming skillfully.
Similarly, Gu et al. also demonstrated the sandwich-structured cellulose composite membranes for achieving the switching of infrared
radiation properties \cite{ref202}. Notably, the unique design can be spread easily to building insulation, energy saving windows and
military equipment.
\begin{figure*}
\centering
\includegraphics[scale=0.45]{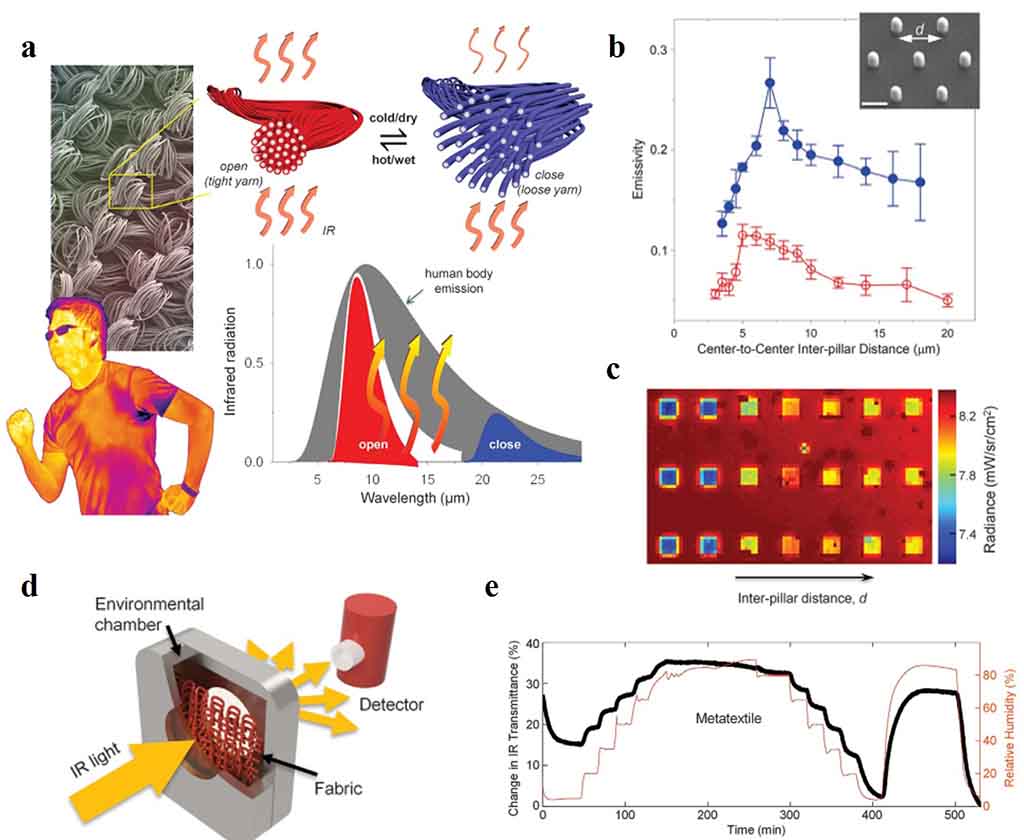}
\caption{Dynamic infrared radiation gating textile. (a) The physical principle of the IR gating textile. (b) Evaluated the connection
between IR emittance and the electromagnetic coupling distance of nanostructure pillar arrays through the measurement of the
environment. The insets is the SEM image of nanostructure pillar arrays in a small area. (c) IR images of the nanostructure pillar
arrays with different spacing on the substrate of 150 $^\circ$C\ . (d) A diagram of the experimental setup for measuring the IR
response of the metafabric versus relative humidity. (e) The comparison of IR transmittance and relative humidity of metatextile vary
with time, which illustrates a favourable dynamic response character between them.\cite{ref98}}
\label{Fig.17*}
\end{figure*}
\subsection{\label{sec5.2}Dynamically switching the MIR emittance}Although the dual-mode textiles above mentioned have enhanced
human body thermal comfort in specific conditions, these textiles are nonresponsive to environmental changes. In recent years,
researchers found that many species in nature have evolved to regulate their own radiation with the environment to satisfy individual
thermal comfort. For instance, Saharan Silver ants have triangular-shaped hairs on the top of the head that rotate in accordance with
the position of the sun, and it can reflect as much sunlight as possible to reduce the energy of absorbing sunlight \cite{ref76}.
Inspired by these animals, some progresses have been developed in the dynamic dual-mode textiles via adjusting the optical
properties of materials.
\begin{figure*}
\centering
\includegraphics[scale=0.45]{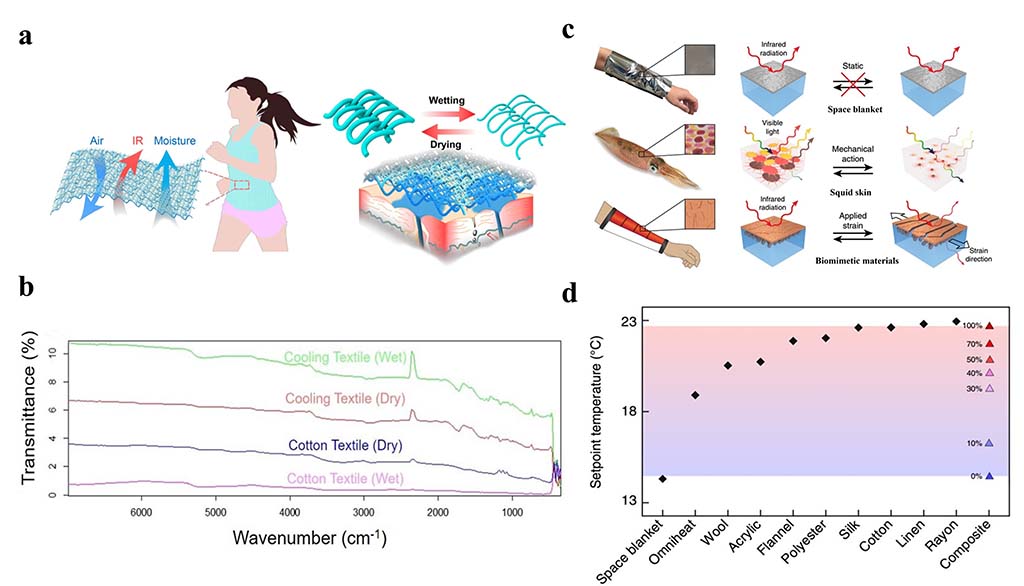}
\caption{Dynamically dual-mode textiles based on adjusting the IR transmittance of materials. (a) A dynamically thermoregulatory
textile that adjusts the infrared transmittance, moisture transport, and air convection of fabric based on the relative humidity of
the skin uses the unique bilayer structure. (b) The infrared  transmittance spectrum of the thermoregulatory textile and cotton
textile in the wet and dry state. \cite{ref223} (c) The tunable human body radiation composite material inspired by squid skin. The
biomimetic materials combining the concept of space blanket with the squid skin achieved the dynamically thermal regulation of body
radiation through the method of mechanical actuation. A space blanket is composed of a solid plastic sheet covered with a dense metal
film, which cannot dynamically adjust the body radiation. The unique squid skin consists of well-organized chromatophore organs
embedded a transparent matrix, which can dynamically regulate body radiation driven by muscle. The thermoregulatory material is
composed of an infrared transparent and soft polymer matrix and well-organized metal domains through the columnar nanostructures
embedded in the matrix. (d) Measured environmental setpoint temperature with satisfying body thermal comfort covered with various
traditional textile samples and the composite textiles at different strains. The different triangles represent various strains.
\cite{ref94}}
\label{Fig.18*}
\end{figure*}

Considering that the body transfers heat to the environment through the form of thermal radiation is essentially related to
electromagnetic waves. Therefore, the dynamic control of human thermal radiation through textiles can be completely from the physical
knowledge related to electromagnetic waves. The electromagnetic spectrum and wave propagation of thermal radiation can be varied by
controlling the distance-dependent electromagnetic interactions between the conductive elements \cite{ref95,ref96,ref97}. Zhang et
al. designed a dramatic textile that dynamically regulates infrared radiation based on tunable electromagnetic wave interactions
\cite{ref98}. Namely, it also modulates the MIR emittance of textiles dynamically, as mentioned above in Eq. (\ref{E6}).
In this design, they fabricated a unique artificial fiber combining a certain amount of conductive material with humidity responsive
material (Fig. \ref{Fig.17*}a).

These yarns made of artificial fibers will become loose reducing the distance of fibers in a relatively hot or humid environment,
causing the electromagnetic resonance coupling. The electromagnetic resonance coupling can shift the emittance of the textile to the
mid-infrared electromagnetic wave to make it more closely matched with the human body thermal radiation, which effectively improves
the heat exchange between the human body and  environment (Fig. \ref{Fig.17*}b-c). Conversely, the yarn will present a state that is
completely opposite to the previous one in a relatively cold or relatively dry environment. It will decrease the heat loss of the
human body achieving a good thermal insulation effect. Besides, experimental results indicate that the thermal radiation of the
carbon-coated nanocolumns is strongly nonlinear dependent on its distance, which proved the feasibility of the novel design. In the
range of 5\%-90\% relative humidity of the human skin, the textiles exhibited an effective infrared response that varied with changes
in humidity (Fig. \ref{Fig.17*}d-e). Obviously, the unique textile can change the radiative heat of the body based on sensing the
relative humidity of the human skin, further satisfying the human thermal comfort in different environments as much as possible.
\begin{figure*}
\centering
\includegraphics[scale=0.45]{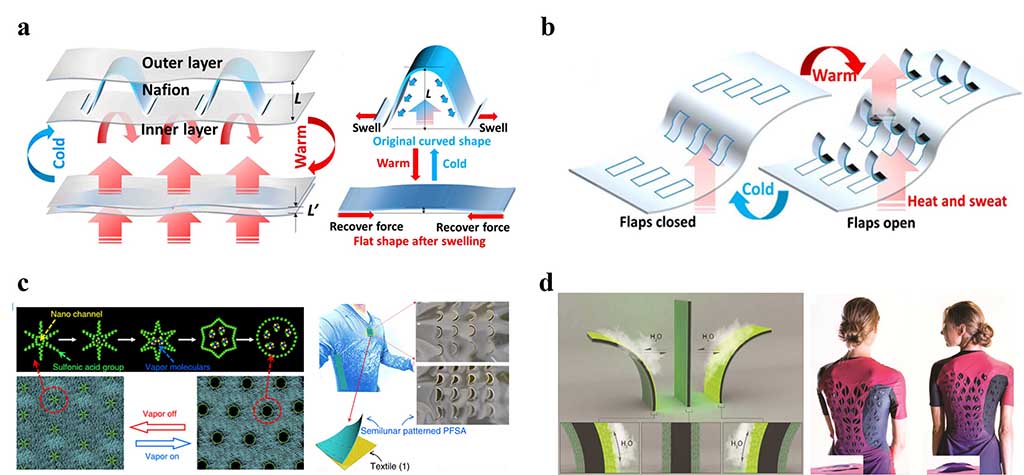}
\caption{Dual-mode thermoregulatory textiles designed by switching the thermal conductance of materials. Smart dual-mode textile (a)
with thickness reversible structure achieved by inserting a Nafion sheet \cite{ref235} and (b) with the flap array structure inspired
by the sweating pore of skin \cite{ref235}, respectively. (c) The PFSA film with nanoscale molecular channels is integrated into a
traditional textile for personal thermal management \cite{ref236}. (d) A biohybrid wearable dual-mode textile utilizing the
hygroscopic property of living cells \cite{ref237}.}
\label{Fig.19*}
\end{figure*}
\subsection{\label{sec5.3}Dynamically switching the MIR transmittance}Relatively, Fu et al. developed a bilayer fabric to adjust
dynamically the infrared transmittance based on the relative humidity change of skin \cite{ref223}. It is also bound up with the MIR transmittance ($\tau_{f}$) of textile in Eq. (\ref{E6}) and (\ref{E9}). The bilayer fabric is composed
of polyethylene terephthalate (PET) yarns inside and moisture-responsive yarns outside (Fig. \ref{Fig.18*}a). The moisture-responsive
yarns can promptly absorb the sweat from the inside PET yarns that maintains the skin in the comfortably dry state. Meanwhile, it
will change its structure from loose to dense, which allows more infrared radiation to the environment achieving an excellent cooling
effect. The dense structure also enlarges the loop size of fabric that improves the heat transfer in the form of air convection. In
contrast, the structure of fabric will be unconsolidated at a lower relative humidity of skin, decreasing infrared radiation
through the fabric achieving a warming effect. At the same time, the loose structure shrinks the loop size of the fabric reducing the
air convection. As shown in Fig. \ref{Fig.18*}b, experimental results indicate that the infrared transmittance of traditional fabric
decreases from dry to wet while the resulting fabric increases from dry to wet, which ensures the human thermal-wet comfort. The
design philosophy of dynamically thermoregulatory textile based on the relative humidity of the skin makes the advanced textile
easier to enter the market.

In addition, the fascinating dynamic color-changing skin of squid, cuttlefish, and octopus have inspired humans to develop
dynamically thermoregulatory textiles based on regulating the infrared transmittance of materials \cite{ref82,ref83,ref84}. In the
case of squid, its skin contains red, yellow, and brown pigment organs, which are composed of inner chromatophore organs surrounded
by innervated muscle cells (Fig. \ref{Fig.18*}c) \cite{ref85,ref87,ref88} . These highly evolved biological structures can dynamically
switch chromatophore organs  between contracted point-like and expanded plate-like states through muscle cells, which in turn can
dynamically change the local coloration and modulate the transmission of light on the skin surface \cite{ref85,ref87,ref88}.

Recently, researchers have combined this unique structure with a high-reflectance space blanket to design a dynamic dual-mode
material \cite{ref94}. The novel composite material consists of a soft, stretchable infrared transparent polymer matrix and the
infrared-reflecting metal domains stack that is stably anchored with the matrix by columnar nanostructures. When the material is not
subject to any mechanical actuation, the infrared-reflecting metal domains with the nanostructure-anchored are densely arranged on
the surface of the polymer, which will strongly reflect the infrared radiation from all directions. However, when mechanical
actuation is applied to the material, the metal domains covering the polymer matrix will be torn apart and the polymer matrix cannot
be completely shielded, which will allow the transmission of the infrared radiation. In essence, the human thermal radiation can be
dynamically regulated by reversible mechanical actuation of the composite material. Specifically, the composite material can be
mechanically driven by external forces to maintain the thermal comfort of the wearer in the ambient temperature range of about 8.2
$^\circ$C\ (Fig. \ref{Fig.18*}d).

\subsection{\label{sec5.4}Dynamically switching the thermal conductance}The dual-mode textiles above mentioned are achieved through
designing the optical properties of textiles from the perspective of regulating the radiative heat dissipation of the human body.
The principle of dynamically switching the heat conduction characters also is applied in textiles to enhance human
thermal comfort. Phase change materials (PCMs) taking advantage of latent heat have been widely embedded into textiles to regulate
dynamically the conductive heat of human body in recent years \cite{ref117,ref120,ref207,ref121,ref211,ref212,ref213,ref110}, and
many reviews have already covered the thermoregulatory textiles with PCMs
\cite{ref20,ref198,ref126,ref18,ref116,ref115,ref113,ref112,ref111}. Hence, we no longer give unnecessary details. Here, we will
concentrate on introducing the dual-mode thermoregulatory textiles that adjust the conductive heat transfer of the human body
utilizing the humidity responsive materials.
\begin{figure*}[htbp]
\makebox[\textwidth][c]{\includegraphics[width=15cm,height=15cm]{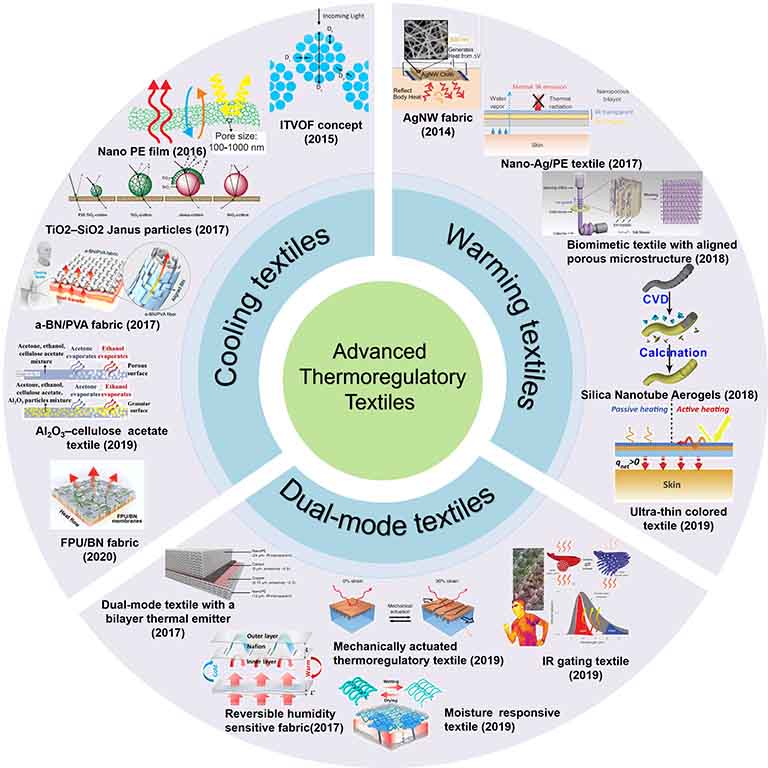}}
\addtolength{\leftskip}{-10pt}
\addtolength{\rightskip}{-10pt}
\caption{Remarkable development of advanced thermoregulatory textiles containing cooling, warming, and dual-mode ones.}
\label{Fig.20*}
\end{figure*}

Interestingly, Zhong et al. demonstrated two kinds of smart thermoregulatory textiles utilizing the thermo-moisture property of
Nafion polymer from DuPont \cite{ref235}. The Nafion film can bend towards the lower humidity side due to the asymmetric swelling of
water-absorbed polymeric chains in Nafion when the opposing faces exist in a humidity difference \cite{ref238}. The first design
tactfully used the Nafion interlayer to actuate the thickness of air layer in textile achieving the dynamic switching of thermal conductance. As aforementioned in Eq. (\ref{E9}), the thickness of air layer is significant association with the effective thermal conductance. As presented in Fig. \ref{Fig.19*}a, the Nafion film is inserted into two fabrics to adjust the air-layer thickness between the upper
and lower layers. As the humidity of the lower layers, Nafion film becomes swell increasing the air-layer thickness, which
accelerates the heat dissipation of skin via enhancing the thermal conductance of the whole textile. Whereas, Nafion film becomes
flat preventing the heat loss of skin by decreasing the thermal conductance of textiles. The second one composing of the Nafion
sheet with the flap array structure mimics the pores of the skin, whose flaps can open or close to regulate the heat transfer of the
human body based on the relative humidity fluctuations (Fig. \ref{Fig.19*}b).

\begin{figure*}
\centering
\includegraphics[scale=0.475]{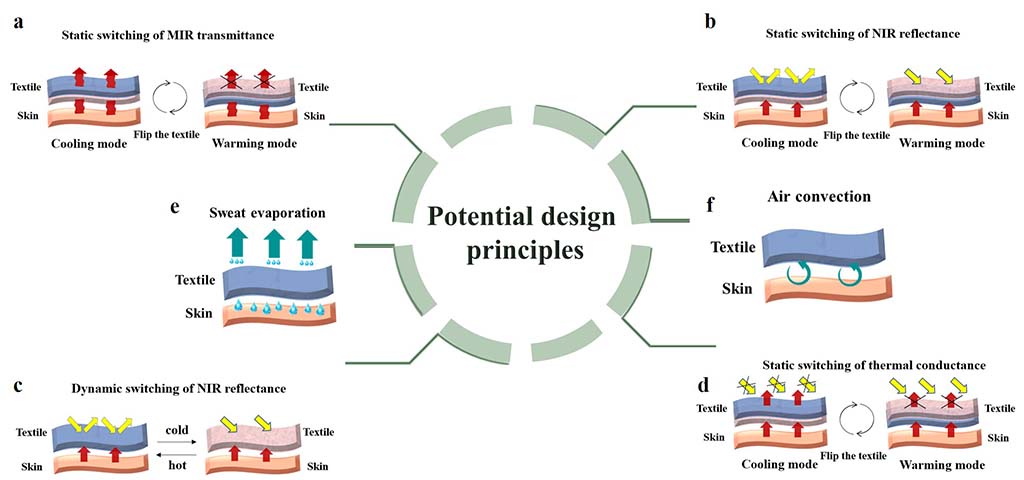}
\caption{The promising perspectives of advanced thermoregulatory textiles designed by adjusting personal heat transfer pathways. In
comparison with the designing principles of single-mode ones, we provide some perspectives during the development of the dual-mode
textiles, including the methods of controlling the two-face MIR transmittance (a), switching the NIR reflectance/absorptance both
statically (b) and dynamically (c) and adjusting the thermal conductance of textiles based on the thermal diode effect (d). Notably,
the impact of  sweat evaporation (e) and thermal convection (f) in the heat transfer pathways of clothed skin should also be explored
carefully in the heat transfer model of human body.}
\label{Fig.21*}
\end{figure*}
Similarly, Mu et al. took advantage of the inherent nanoscale molecular channels of perfluorosulfonic acid ionomer (PFSA) film
fabricating a kirigami-inspired humidity-driven thermoregulatory textile (Fig. \ref{Fig.19*}c) \cite{ref236}. The unique nanoscale
molecular channels in PFSA film improve the mass transfer of vapor that greatly shortens response time compared with the microporous
structure. Besides, the hygroscopic property of living cells also provides us a new insight into the development of smart
thermoregulatory textiles. Inspired by the concept, Wang et al. fabricated the biohybrid films that can change its shape reversibly
to adjust the heat dissipation of skin based on the environmental humidity gradients (Fig. \ref{Fig.19*}d) \cite{ref237}. The bending
effect of biohybrid film is obtained by the net contractile force generated via all of the cells. These ingenious design methods to
the dual-mode thermoregulatory textiles bring the broad prospect.

\begin{figure*}
\centering
\hspace{-1cm}
\includegraphics[scale=0.5]{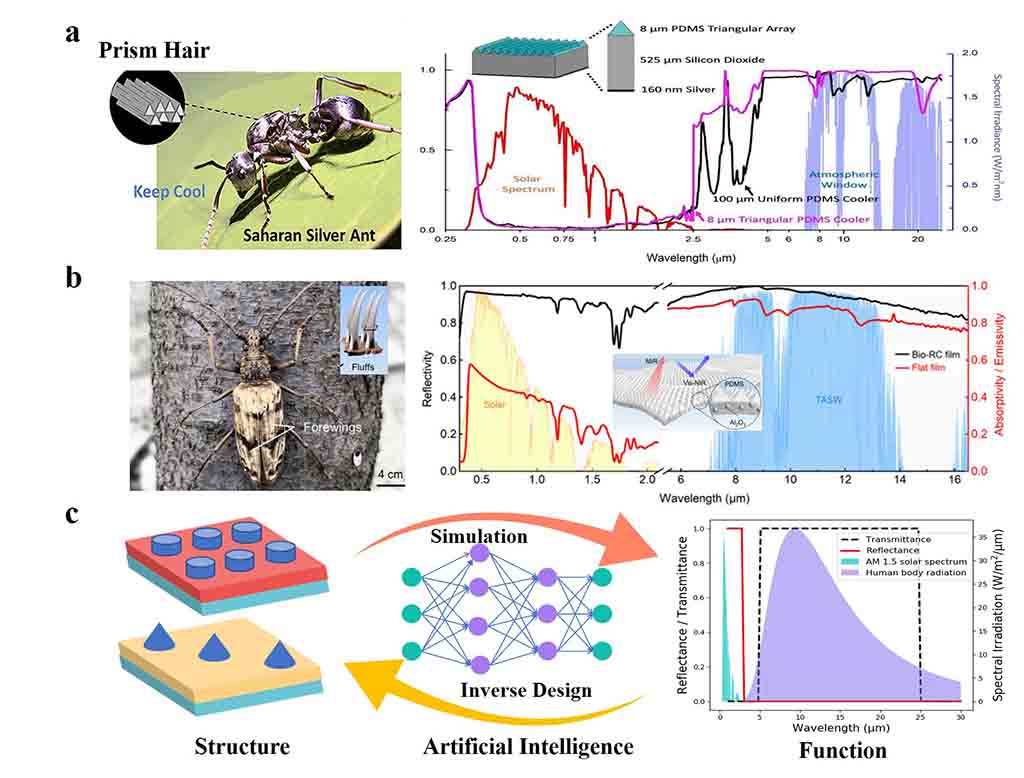}
\caption{The potential designing alternatives of advanced thermoregulatory textiles. (a) A radiative cooling device with the
spectrally selective property achieved by mimicking the unique triangular prism structure of the Sahara silver ants' hair
\cite{ref239}. (b) A flexible radiative cooling film taking inspiration from the longicorn beetles' dual-scale fluffs \cite{ref240}.
These attractive structures of natural creatures indeed provide a new horizon of expectation for developing the fiber structure with
the excellent thermal regulating properties. (c) Inverse engineering the target thermo-textiles by artificial intelligence has the
potential to create some novel structures beyond our imagination.}
\label{Fig.22*}
\end{figure*}

\section{\label{sec6}Summary, challenges, and outlooks}
As a summary, we have reviewed the current studies of advanced thermoregulatory textiles with elaborated rich heat transfer pathways, the remarkable development of which are categorized as cooling, warming, and dual-mode as shown in Fig. \ref{Fig.20*}. On the one hand, we have introduced the heat exchange pathways between the human body and the external environment. On the other
hand, we have categorized advanced thermoregulatory textiles based on multiple heat exchange pathways of thermal radiation and non-radiative conduction between the human body and environment. 
It indicates that the novel textiles with thermal regulation function are promising to achieve commercial products, improving the human body's ability to withstand the harsher living environment. The rapid development of the advanced thermoregulatory textiles is not only a historic change in the textile industry, but also a great progress forward in the sustainable development of energy. In addition, the energy crisis and rapid climate warming have hindered the sustainable development of society. Advanced textiles focus on adjusting the personal micro-environment to improve the human thermal
comfort in different environments. Both theoretical analysis and experimental works show that the novel designs have potential to regulate human body heat transfer, which have a broad prospect in superseding air conditioners, electric heaters, and warming equipment.

Although the novel thermoregulatory textiles have made rapid progress, there are still many challenges in the commercial road. Firstly, the nanostructures is commonly used for the materials in thermoregulatory textiles metioned above. The harm of nanomaterials
between body and environment has still not been precisely investigated. Compared with ordinary pollutants, nanomaterials may be harmful to human beings. The reason is that nanomaterials have a smaller size than micron particles, making
nanoparticles enter easily the human body. In addition, the thermoregulating textiles composed of the conductive materials have external
safety issues in the commercialization of these advanced textiles. Secondly, the fabrication of the thermoregulatory
textiles are complex, difficult, and expensive. Meanwhile, the wearing comfort of the advanced textiles can be further improved in
the future. Lastly, the impact of the most basic dyeing technology on the thermal regulation of advanced textiles also requires further exploration. In
a word, the development of these thermoregulatory textiles still faces enormous challenges. Hence, understanding the designed principles of textiles with thermal regulation functions and rich microstructure-function correspondences can inspire more researchers to find feasible ways to overcome difficulties, thus accelerating the commercialization of new textiles.

Meanwhile, there are still plenty of gaps to achieve advanced thermoregulatory textiles, for which we would like to put forward some outlooks to the further development in the future:

{\bf (1) Missing blocks for adjusting heat pathways.} As presented in Fig. \ref{Fig.3*}, we summarized the potential design principles
of currently advanced thermoregulatory textiles based on the adjustment of heat transfer pathways. There are yet many research gaps that are not well discussed in literature. For example, the dual-mode
thermoregulatory textiles can be achieved via switching the two-face MIR transmittance (Fig. \ref{Fig.21*}a) and thermal conductance of textiles (Fig. \ref{Fig.21*}d) based on the thermal diode effect \cite{ref226,ref230,ref231,ref232},
and controlling the NIR reflectance/absorptance both statically (Fig. \ref{Fig.21*}b) and dynamically (Fig. \ref{Fig.21*}c) inspired by the design ideas of the single-mode ones, which can further broaden one's horizon in the development of the dual-mode advanced thermoregulatory textiles.

{\bf (2) Impact of evaporation and convection.} In addition to the radiation and conduction that the present review focused on, sweat
evaporation (Fig. \ref{Fig.21*}e) and air convection (Fig. \ref{Fig.21*}f) are also regarded as important roles in the human body heat
transfer pathways. However, most of the advanced thermoregulatory textiles above mentioned ignored the impact of these heat
transfer pathways. Although few works have tried to develop thermoregulatory textiles based on regulating the sweat evaporation
\cite{ref243,ref256}, it remains challenge to tackle and solve the heat management problems caused by air convection and sweat evaporation both in theory and practice.

{\bf (3) Biologically inspired design of microstructures.} The biological world provides many species with an astonishing
thermal regulating ability for addressing severe weather conditions. For instance, an increasing number of warming textiles are
demonstrated in recent years inspired by the porous structure of polar bear hair \cite{ref77,ref189}.
Moreover, some natural creatures living in the tropical regions, silvery Saharan ants, Goliathus goliatus and Neocerambyx gigas, etc
have been discovered to have fascinating thermal control behaviors with high NIR reflectance and MIR emittance due to their unique shaped hairs (Fig. \ref{Fig.22*}a,b) \cite{ref239,ref240}.
Intriguingly, the dual-mode switching between cooling and warming also have been proposed based on the fascinating dynamically color-changing ability of squid skin \cite{ref94,ref260}.
As a result, the  unique thermal regulating ability of these natural creatures should be investigated carefully to provide new insights into the development of bio-inspired advanced thermoregulatory textiles.

{\bf (4) Artificial intelligence inverse design of textiles.} Beside the bio-inspired design of textiles for enhancing human thermal
comfort, artificial intelligence assisted inverse design can alternatively guide the design of novel advanced textile microstructures that could
over-perform beyond natural structures. The methods of various machine learnings can be exploited to obtain some desired composite
structures according to the target properties, like radiation spectra \cite{ref241,ref242}. This scheme can further provide new visions to explore more effective ways to enhance the abilities of personal thermal regulation of the advanced textiles (Fig. \ref{Fig.22*}c).

\begin{acknowledgments}{This work is supported by the National Natural Science Foundation of China (No. 11935010, No. 11775159, No. 51925302, and No. 51673037), the Natural Science Foundation of Shanghai (No. 18ZR1442800 and No. 18JC1410900) and the Opening Project of Shanghai Key Laboratory of Special Artificial Microstructure Materials and Technology.}
\end{acknowledgments}


\section*{References}
\nocite{*}
\bibliography{ref}

\end{document}